\documentclass[reprint,superscriptaddress,
frontmatterverbose,showpacs,showkeys,preprintnumbers,amsmath,amssymb,aps,pra,floatfix]{revtex4-1}

\usepackage[pagebackref=false,colorlinks,linkcolor=red,citecolor=magenta]{hyperref}
\usepackage{graphicx}
\begin{document}

% Use the \preprint command to place your local institutional report
% number in the upper righthand corner of the title page in preprint mode.
% Multiple \preprint commands are allowed.
% Use the 'preprintnumbers' class option to override journal defaults
% to display numbers if necessary
%\preprint{}

%Title of paper
\title{ Force sensing based on coherent quantum noise cancellation in a hybrid optomechanical cavity with squeezed-vacuum injection}
\author{Ali Motazedifard}
\email{motazedifard.ali@gmail.com}
\author{F. Bemani}
\email{foroudbemani@gmail.com}
\address{Department of Physics, Faculty of Science, University of Isfahan, Hezar Jerib, 81746-73441, Isfahan, Iran}
\author{M. H. Naderi}
\email{mhnaderi@sci.ui.ac.ir}
\author{R. Roknizadeh}
\email{rokni@sci.ui.ac.ir}
\address{Quantum Optics Group, Department of Physics, Faculty of Science, University of Isfahan, Hezar Jerib, 81746-73441, Isfahan, Iran}
\author{D. Vitali}
\email{david.vitali@unicam.it}
\address{Physics Division, School of Science and Technology, University of
Camerino, I-62032 Camerino, Italy, and INFN, Sezione di Perugia, Perugia, Italy}
\date{\today}
\begin{abstract}
We propose and analyse a feasible experimental scheme for a quantum force sensor based on the elimination of back-action noise through coherent quantum noise cancellation (CQNC) in a hybrid atom-cavity optomechanical setup assisted with squeezed vacuum injection. The force detector, which allows for a continuous, broad-band detection of weak forces well below the standard quantum limit (SQL), is formed by a single optical cavity simultaneously coupled to a mechanical oscillator and to an ensemble of ultracold atoms. The latter acts as a negative-mass oscillator so that atomic noise cancels exactly the back-action noise from the mechanical oscillator due to destructive quantum interference. Squeezed vacuum injection enforces this cancellation and allows to reach sub-SQL sensitivity in a very wide frequency band, and at much lower input laser powers.
%Moreover, squeezed injection suppresses the low frequency noise by factor of $1/4N$.
\end{abstract}
\pacs{42.50.Dv, 03.65.Ta, 42.50.Wk}
\keywords{Force sensing, Standard quantum limit, Hybrid optomechanics, Squeezed vacuum state }

%\maketitle must follow title, authors, abstract, \pacs, and \keywords
\maketitle

% body of paper here - Use proper section commands
% References should be done using the \cite, \ref, and \label commands
\section{Introduction}
Every measurement is affected by noise, degrading the signal and consequently reducing the accuracy of the measurement. However, noise cancellation techniques can be applied if the noise can be identified and measured separately, as, for example, in the acoustic domain \cite{1}. The application of noise cancellation to quantum systems has recently been introduced \cite{PRL,PRX} by using the so-called \textit{coherent quantum noise cancellation} (CQNC) scheme, which relies on quantum interference. The basic idea is that under certain conditions, it is possible to introduce an ``anti-noise" path in the dynamics of the system which can be employed to cancel the original noise path via destructive interference.

The measurement of weak forces at the quantum limit \cite{4} and the search for quantum behavior in macroscopic degrees of freedom have been some of the motivations at the basis of the development of cavity optomechanics \cite{5,6,7}. In a force measurement based on an optomechanical scheme \cite{8,9}, the competition between shot noise and radiation pressure back-action noise leads to the notion of SQL \cite{4}. Shot noise is a known effect limiting high-precision interferometry at high frequencies \cite{10}, while radiation pressure noise, recently observed for the first time \cite{11,12}, becomes relevant only at large enough powers and will be limiting in the low-frequency regime next-generation gravitational-wave detectors \cite{13}. These two noise sources have opposite scaling with the input field power: increasing the input power in order to enhance the measurement strength and decrease the shot noise leads to increase the measurement back-action noise. Therefore, in order to improve the force detection sensitivity one has to eliminate the backaction noise.

There are various proposals for reducing quantum noise and overcoming the SQL in force measurements, including frequency-dependent squeezing of the input field \cite{14}, variational measurements \cite{15,16}, the use of Kerr medium in a cavity \cite{17}, a dual mechanical oscillator setup \cite{18}, the optical spring effect \cite{19}, and two-tone measurements \cite{20,21,22,23}. Preliminary experimental demonstrations of these ideas have been already carried out~\cite{24,25,26,27,28,pontin}, and recent clear demonstrations of quantum-nondemolition measurements have been given in~\cite{22,teufel}.

A different approach for sub-SQL measurements has recently been introduced \cite{PRL,PRX}, based on the CQNC of back action noise via quantum interference. The idea is based on introducing an``anti-noise" path in the dynamics of the optomechanical system via the addition of an ancillary oscillator which manifests an equal and opposite response to the light field, i.e, an oscillator with an effective negative mass. In the context of atomic spin measurements an analogous idea for coherent backaction cancellation was proposed independently \cite{29,30}, and has been applied for magnetometry below the SQL \cite{31}, demonstrating that Einstein-Podolski-Rosen (EPR)-like entanglement of atoms generated by a measurement enhances the sensitivity to pulsed magnetic fields.
The original proposal \cite{PRL} focused on the use of an ancillary cavity that is red-detuned from the optomechanical cavity. A quantum non-demolition coupling of the electromagnetic fields within the two cavities yields the necessary anti-noise path, so that the backaction noise is coherently cancelled. Ref. \cite{maximilian} considered in more detail the all-optical realization of the CQNC proposal put forwarded in \cite{PRL,PRX}, and found that the requirements for its experimental implementation appear to be very challenging, especially for the experimentally relevant case of low mechanical frequencies and high-quality mechanical oscillators (MO) such as gravitational wave detectors. Other setups, which provide effective negative masses of ancillary systems for CQNC, have been suggested based on employing Bose-Einstein condensates \cite{34}, or the combination of a two-tone drive technique and positive-negative mass oscillators \cite{35}.

In recent years, hybrid optomechanical systems assisted by the additional coupling of the cavity mode with an atomic gas have attracted considerable attention. It has been found that
%these systems exhibit advantages in many aspects. In particular, by inducing an additional nonlinear effect,
the additional atomic ensemble may lead to the improvement of optomechanical cooling \cite{36,37,38,39,40}, thereby providing the possibility of ground state cooling outside the resolved sideband regime \cite{41,42}. Moreover, the coupling of the mechanical oscillator to an
atomic ensemble can be used to generate a squeezed state of the mechanical mode \cite{Nori}, or robust EPR-type entanglement between collective spin variables of the atomic medium and the mechanical oscillator \cite{29,genes}.

Inspired by the above considerations, more recently a theoretical scheme for CQNC based on a dual cavity atom-based optomechanical system has been proposed \cite{meystre}. In this scheme, a MO used for force sensing is coupled to an ultracold atomic ensemble trapped in a separate optical cavity which behaves effectively as an effective negative mass oscillator (NMO). The two cavities are coupled via an optical fiber.
This system is a modification of the setup suggested for hybrid cooling and electromagnetically induced transparency \cite{meystrecooling} and the interaction between the optomechanical cavity and the atomic ensemble leads to the CQNC. The atomic ensemble acts as a more flexible NMO, for which the ``impedance matching'' condition of a decay rate identical to the mechanical damping rate is easier to satisfy with respect to the full-optical implementation of Ref.~\cite{maximilian}.

Here we propose to simplify and improve the atomic ensemble implementation of CQNC of Ref.~\cite{meystre} by considering a different setup, involving only a \emph{single} optomechanical cavity and a single cavity mode, coupled also to an atomic ensemble, which is also injected by squeezed vacuum (see Fig.\ref{fig1} (a)).
The atomic ensemble is coupled to the radiation pressure and the coupling strength of the atom-field interaction is modulated. We show that the interaction between the optomechanical cavity and the atomic ensemble leads to an effective NMO that can provide CQNC conditions able to eliminate the backaction noise of the MO. In fact, destructive quantum interference between the collective atomic noise and the backaction noise of the MO realizes an `anti-noise' path, so that the backaction noise can be cancelled (Fig.\ref{fig1} (b)). CQNC conditions are realized when the optomechanical coupling strength and the mechanical frequency are equal to the coupling strength of the atom-field interaction and to the effective atomic transition rate, respectively. Furthermore, the dissipation rate of the MO needs to be matched to the decoherence rate of the atomic ensemble.

Here we exploit the injection of appropriately squeezed vacuum light in order to control and improve the noise reduction for force detection, applying within this new scenario, the properties of squeezing. In fact, it is well known that the injection of a squeezed state in the unused port of a Michelson interferometer can improve interferometric measurements \cite{caves80,caves,Reynaud,pace,McKenzie,QuantumMeasurement,Chen}, as recently demonstrated in the case of gravitational wave interferometers~\cite{LIGO}.
The improvement of the performance of measurement via squeezing injection has also been demonstrated in other interferometers, such as the Mach-Zehnder \cite{Mach-Zehnder}, Sagnac \cite{Sagnac}, and polarization interferometers \cite{polarization}. Squeezing-enhanced measurement have been realized also within optomechanical setups: an experimental demonstration of squeezed-light enhanced mechanical transduction sensitivity in microcavity optomechanics has been reported in \cite{transduction}. Moreover, by utilizing optical phase tracking and quantum smoothing techniques, improvement in the detection of optomechanical motion and force measurements with phase-squeezed state injection has also been verified experimentally \cite{phase-squeezed state}. Finally the improvement in position detection by the injection of squeezed light has been recently demonstrated also in the microwave domain in Ref.~\cite{Clark}. We also notice that it has been recently theoretically shown that even the intracavity squeezing generated by parametric down conversion can enhance quantum-limited optomechanical position detection through de-amplification~\cite{intracavity squeezing}. More recently Ref. \cite{Lotfipor} has investigated the response of a mechanical oscillator in an optomechanical cavity driven by a squeezed vacuum and has shown when it can be used as a high sensitive nonclassical light sensor.

In the present paper we show that if the cavity mode is injected with squeezed light with an appropriate phase, back-action noise cancellation provided by CQNC is much more effective because squeezing allows to suppress the shot noise contribution at a much smaller input power, and one has a significant reduction of the force noise spectrum even with moderate values of squeezing and input laser power.

The paper is organized as follows. Section \ref{sec2} is devoted to the description of the model. The linear quantum Langevin equations of motion for the dynamical variables involved in the sensing process are derived in Section \ref{sec3}. The main results for force sensing and the increased sensitivity achieved in the case of back-action cancellation provided by CQNC are given in Section \ref{sec4}. Finally, the conclusions are summarized in Section \ref{sec5}.

\section{\label{sec2}The system }

\begin{figure}
\begin{center}
\includegraphics[width=8.7cm]{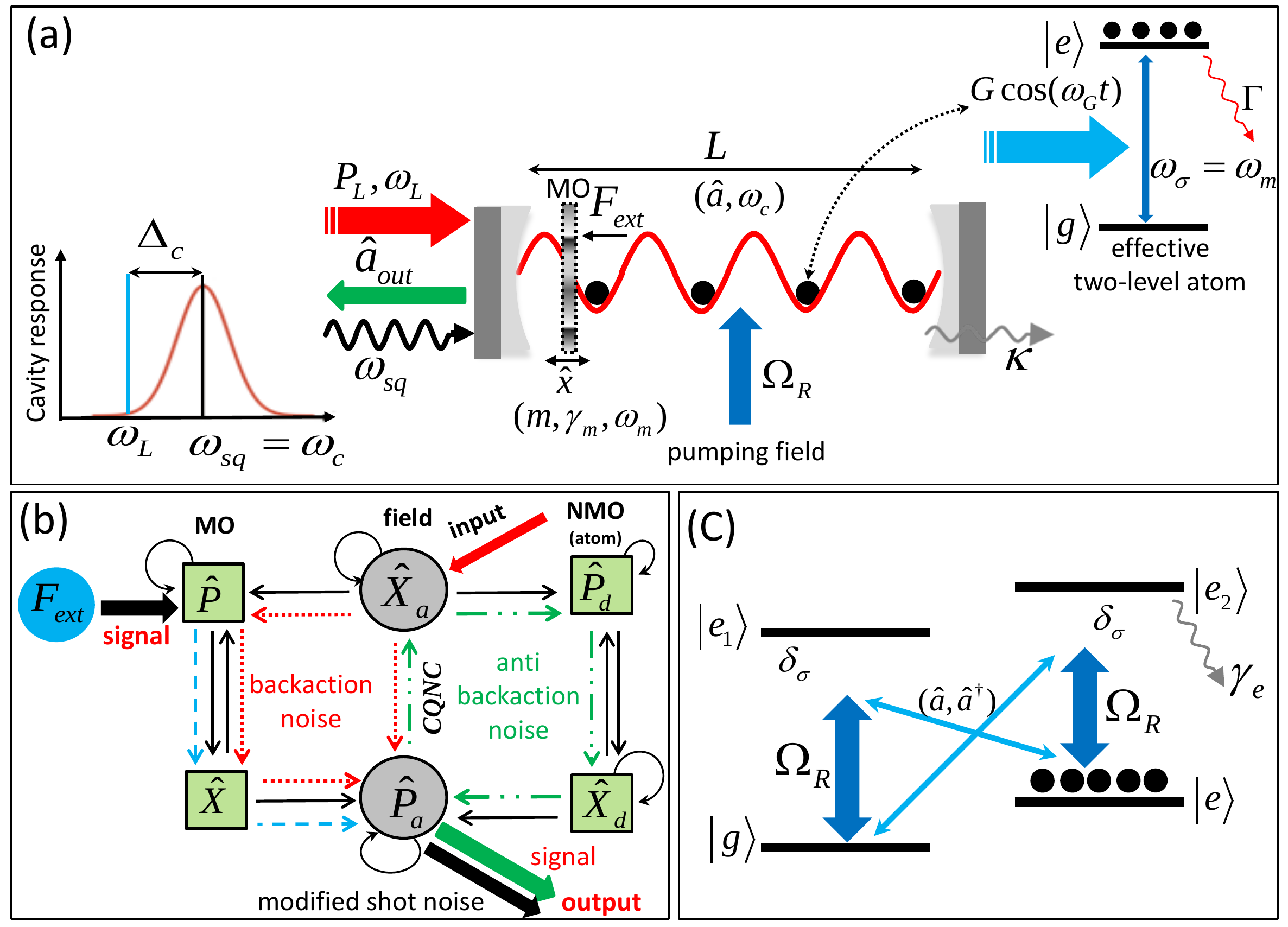}
\end{center}
\caption{(Color online) (a) Schematic description of the system under consideration. A mechanical oscillator with frequency $\omega_m$ is placed within a single-mode Fabry-P\'{e}rot cavity containing an atomic ensemble that can be controlled by a classical pumping field with Rabi frequency $\Omega_R$ with effective transition rate $\omega_\sigma = \omega_m$. An external force $F_{ext}$ is exerted on the mechanical oscillator acting as a sensor. The cavity is driven by a classical laser field with power $P_L$ and frequency $\omega_L$, and also a squeezed light field, resonant with the cavity mode, $\omega_{sq}=\omega_c$ , is injected into the cavity.
(b) Flow chart representation of the backaction noise cancellation caused by the anti-noise path associated with the interaction of the cavity mode with the atomic ensemble acting as a negative mass oscillator (NMO).
(c) Atomic scheme leading to the effective Faraday interaction, with a double $\Lambda$ atomic system coupled to the intracavity mode $\hat a$ (thin blue line) and driven by a classical control field (thick blue line) of frequency $\omega_G=\omega_c$ resonant with the cavity mode. }
\label{fig1}
\end{figure}

The optomechanical setup considered in this paper is schematically described in Fig.~\ref{fig1}(a). The system consists of a single Fabry-P\'{e}rot cavity in which a MO, serving as a test mass for force sensing, is directly coupled to the radiation pressure of an optical cavity field. Furthermore, the cavity contains an ensemble of effective two-level atoms that is coupled to the intracavity mode.
As is shown in Fig.~\ref{fig1}(c), and will be detailed further below, the two-level atomic ensemble with time-modulated coupling constant considered in this scheme is achievable by considering a double $\Lambda$-type atomic ensemble driven by the intracavity light field and by a classical control field.

We consider a standard optomechanical setup with a single cavity mode driven by a classical laser field with frequency $\omega_L$, input power $P_L$, interacting with a single mechanical mode treated as a quantum mechanical harmonic oscillator with effective mass $m$, frequency $\omega_m$, and canonical coordinates $\hat x$ and $\hat p$, with $[\hat x, \hat p]=i\hbar $. This single mode description can be applied whenever scattering of photons from the driven mode into other cavity modes is negligible~\cite{53}, and if the detection bandwidth is chosen such that it includes only a single, isolated, mechanical resonance and mode-mode coupling is negligible~\cite{54}. Moreover, the cavity is injected by a squeezed vacuum field with central frequency $\omega_{sq}$ which is assumed to be resonant with the cavity mode $\omega_{sq}=\omega _c$.

The total Hamiltonian describing the system is given by
\begin{equation}
\hat H = {\hat H_{c}} + {\hat H_m} + {\hat H_{om}} + {\hat H_{d}} + {\hat H_{at}} + {\hat H_{F}},
\label{E1}
\end{equation}
where $\hat H_{c}$ describes the cavity field, $\hat H_m$ represents the MO in the absence of the external force $F_{ext}$, $\hat H_{om}$ denotes the optomechanical coupling, $\hat H_{d}$ accounts for the driving field, $\hat H_{at}$ contains the atomic dynamics, and $\hat H_F$  denotes the contribution of the external force. The first four terms in the Hamiltonian of Eq. (\ref{E1}) are given by
\begin{subequations}\label{E2}
\begin{eqnarray}
&&{{\hat H}_c} = \hbar {\omega _c}{{\hat a}^\dag }\hat a ,\\
&&{{\hat H}_m} = \hbar {\omega _m}{{\hat b}^\dag }\hat b = \frac{{{{\hat p}^2}}}{{2m}} + \frac{1}{2}m\omega _m^2{{\hat x}^2},\\
&&{{\hat H}_{om}} =  \hbar {g_0}{{\hat a}^\dag }\hat a(\hat b + {{\hat b}^\dag }),\\
&&{{\hat H}_d} = i\hbar {E_L}({{\hat a}^\dag }{e^{ - i{\omega _L}t}} - \hat a{e^{i{\omega _L}t}}),
\end{eqnarray}
\end{subequations}
where $\hat a$ and $\hat b$ are the annihilation operators of the cavity field and the MO, respectively, whose only nonzero commutators are $[ \hat a,\hat a^{\dag}] =[ \hat b,\hat b^{\dag}]=1$. Furthermore, $\hat x = { x_{ZPF}}(\hat b + {{\hat b}^\dag })$ and $\hat p = i{ p_{ZPF}}({{\hat b}^\dag } - \hat b)$, with ${ x_{ZPF}} = \sqrt {\hbar /2m{\omega _m}}$ and ${p_{ZPF}} = \hbar /2{ x_{ZPF}}$ the zero-point position and momentum fluctuations of the MO. ${g_0} = (d\omega _c/dx){ x_{ZPF}}$ is the single-photon optomechanical strength, while $E_L = \sqrt{P_L \kappa_{in}/\hbar \omega_L}$, with $\kappa_{in}$ the coupling rate of the input port of the cavity.

For the atomic part, we consider an ensemble of $N$ ultracold four-level atoms interacting non-resonantly with the intracavity field and with a classical control field with Rabi frequency $\Omega_R$ and frequency $\omega
_G$ (see Fig. \ref{fig1}(c)). Considering the far off-resonant interaction, the two excited states $\left| {{e_1}} \right\rangle$ and $\left| {{e_2}} \right\rangle$ will be only very weakly populated. In this limit, these off-resonant excited states can be adiabatically eliminated so that the light-atom interaction reduces the coupled double - $\Lambda$ system to an effective two-level system, with upper level $\left| {{e}} \right\rangle$ and lower level  $\left| {{g}} \right\rangle$, (Fig. \ref{fig1}(c)), driven by the so-called Faraday or quantum non-demolition interaction \cite{Quantum interface between light and atomic ensembles}. Apart from the light-matter interface, the Faraday interaction has important applications also in continuous non-demolition measurement of atomic spin ensembles \cite{Atomic2}, quantum-state control/tomography \cite{tomography} and magnetometry \cite{magnetometry}. In the system under consideration, we also assume that a static external magnetic field tunes the Zeeman splitting between the states $\left| {{e}} \right\rangle$ and $\left| {{g}} \right\rangle$ into resonance with the frequency $\omega_m$ of the MO.

Considering the effective two-level model for the atomic ensemble, we introduce the collective spin operators
\begin{subequations}\label{E3}
\begin{eqnarray}
&&{{\hat S }_ + } = \sum_{i = 1}^N {\left| e^{(i)} \right\rangle } \left\langle g^{(i)} \right| = {({{\hat {S} }_ - })^\dag }  ,\\
&&{{\hat S }_z} =  \frac{1}{{2 }}\sum_{i = 1}^N {\left| e^{(i)} \right\rangle } \left\langle e^{(i)} \right| - \left| g^{(i)}  \right\rangle \left\langle g^{(i)}  \right| ,
\end{eqnarray}
\end{subequations}
where \textit{i} labels the different atoms. The collective spin operators obey the commutation relations $\left[ {{{\hat S }_ + },{{\hat S }_ - }} \right] = 2{{\hat S }_z}$ and $\left[ {{{\hat S }_ \mp },{{\hat S }_z}} \right] =  \pm {{\hat S }_z}$, so that the effective Hamiltonian of the atomic ensemble can be written as
\begin{equation}
{{\hat H}_{at}} = \hbar {\omega _m }{{\hat S }_z} + \hbar G_0\cos ({\omega _{^G}}t)(\hat a + {{\hat a}^\dag })({{\hat S }_ + } + {{\hat S }_ - }),
\label{E4}
\end{equation}
where $G_0 = {E_0}({\Omega _R}/{\delta _\sigma})$ is the atom-field coupling, with $E_0$ and $\delta_{\sigma}$ denoting the cavity-mode Rabi frequency and the detuning of the control beam from the excited atomic states, respectively. Now we assume that the atoms are initially pumped in the hyperfine level of higher energy, $\left| {{e}} \right\rangle$, which results in an inverted ensemble that can be approximated for large $N$ by a harmonic oscillator of negative effective mass. This fact can be seen formally using the Holstein-Primakoff mapping of angular momentum operators onto bosonic operators~\cite{HP}. In our case we have a total spin equal to $N/2$ and one can introduce an effective atomic bosonic annihilation operator $\hat{d}$ such that $\hat S_z= N/2-\hat{d}^{\dag} \hat{d}$, $\hat S_+ = \sqrt{N}\left[1-\hat{d}^{\dag} \hat{d}/N\right]^{1/2} \hat{d}$, $\hat S_- = \sqrt{N}\hat{d}^{\dag}\left[1-\hat{d}^{\dag} \hat{d}/N\right]^{1/2} $, so that the commutation rules are preserved. As long as the ensemble remains close to its fully inverted state, we can take $\hat{d}^{\dag} \hat{d}/N \ll 1$ and approximate $\hat S_- \simeq \sqrt{N}\hat{d}^{\dag}$, $\hat S_+ \simeq \sqrt{N}\hat{d}$. Therefore, under the bosonization approximation, we can rewrite Eq.~(\ref{E4}) as
\begin{equation}
{{\hat H}_{at}} =  - \hbar {\omega _m}\hat{d}^{\dag} \hat{d} + \hbar G\cos ({\omega _{^G}}t)(\hat a + {{\hat a}^\dag })({{\hat{d}+\hat{d}^{\dag} }}),
\label{E5}
\end{equation}
which shows that the atomic ensemble can be effectively treated as a NMO, coupled with the collective coupling $G=G_0\sqrt{N}$ with the cavity mode.
Moving to the frame rotating at laser frequency $\omega_L$, where $\hat a \to \hat a e^{-i\omega_L t}$, choosing the resonance condition $\omega_G=\omega_L$, and applying the rotating wave approximation in order to neglect the fast rotating terms, i.e., the terms proportional to $e^{\pm i(\omega_G+\omega_L)t}$, one gets
\begin{equation}
{{\hat H}_{at}}=  - \hbar {\omega _m}\hat{d}^{\dag} \hat{d} + \hbar \frac{G}{2}(\hat a + {{\hat a}^\dag })({{\hat{d}+\hat{d}^{\dag} }}).
\label{E6}
\end{equation}
Therefore, the total Hamiltonian of the system in the frame rotating at laser frequency $\omega_L$ is time-independent and can be written as
\begin{eqnarray}
&&\hat H = \hbar {\Delta_{c0}}{{\hat a}^\dag }\hat a + \hbar {\omega _m}{{\hat b}^\dag }\hat b - \hbar {\omega _m}\hat{d}^{\dag} \hat{d} + \hbar {g_0}{{\hat a}^\dag }\hat a(\hat b + {{\hat b}^\dag }) \nonumber \\
&&\qquad + \hbar \frac{G}{2}(\hat a + {{\hat a}^\dag })({{\hat{d}+\hat{d}^{\dag} }}) + i\hbar {E_L}({{\hat a}^\dag } - \hat a) ,
\label{E7}
\end{eqnarray}
where $\Delta_{c0}=\omega_c-\omega_L$.

\section{\label{sec3}dynamics of the system}

The dynamics of the system is determined by the quantum Langevin equations obtained by adding damping and noise terms to the Heisenberg equations associated with the Hamiltonian of Eq.~(\ref{E7})~\cite{vitali noise membrane},
\begin{subequations}\label{E8}
\begin{eqnarray}
&& \dot {\hat x} = \hat{p}/m\\
&& \dot {\hat p} =  - m{\omega _m^2}\hat x -2 p_{ZPF} g_0{{\hat a}^\dag }\hat a - \gamma _m \hat p +\eta +\tilde{F}_{\rm ext} ,\\
&& \dot {\hat a} =  - i{\Delta_{c0}}\hat a - i g_0 \hat a \frac{\hat{x}}{ x_{ZPF}}  - i\frac{G}{2}({{\hat{d}+\hat{d}^{\dag} }}) + E_L \nonumber \\
&&\qquad - \frac{\kappa }{2}\hat a + \sqrt \kappa  {{\hat a}^{in}} ,\\
&&\dot{\hat d} = i{\omega _m }\hat d - i\frac{G}{2}(\hat a + {{\hat a}^\dag }) - \frac{\Gamma }{2}\hat d + \sqrt \Gamma  \hat d^{in} ,
\end{eqnarray}
\end{subequations}
where $\gamma_m$ is the mechanical damping rate, $\Gamma$ is the collective atomic dephasing rate, $\kappa$ denotes the cavity photon decay rate. We have also considered an external classical force $\tilde{F}_{\rm ext}$ which has to be detected by the MO. The system is also affected by three noise operators: the thermal noise acting on the MO, $\eta(t) $, the optical input vacuum noise, ${{\hat a}^{in}}$, and the bosonic operator describing the optical vacuum fluctuations affecting the atomic transition, $\hat d^{in}$ \cite{GZ}. These noises are uncorrelated, and their only nonvanishing correlation functions are $\langle {{\hat a}^{in}}(t) {{\hat a}^{in}}(t)^{\dag}\rangle = \langle {{\hat d}^{in}}(t) {{\hat d}^{in}}(t)^{\dag}\rangle = \delta (t - t')$ \cite{GZ}.  Here, we have assumed that the external classical force has no quantum noise.
The Brownian thermal noise operator $\eta(t)$ obeys the following correlation function \cite{vitali noise membrane}
\begin{equation}
\left\langle {\eta(t)\eta (t')} \right\rangle \!  =\! m \! \gamma_m \! \hbar \! \int \! {\frac{{d\omega }}{{2\pi }}} \omega {e^{ - i\omega (t - t')}}\! \left[\! \coth\left(\frac{\hbar \omega}{2 k_B T}\right) \! +\! 1 \right] \! ,
\label{E15}
 \end{equation}
where $T$ is the temperature of the thermal bath of the MO. The mechanical quality factor $Q_m=\omega_m/\gamma_m $ is typically very large, justifying the weak damping limit where the Brownian noise can be treated as a Markovian noise, with correlation function \cite{vitali noise membrane}
\begin{equation}
\left\langle {\eta (t) \eta (t')} \right\rangle \!  \simeq \!  \hbar  m \! \gamma_m \!   \left[ \omega_m (2{{\bar n}_m} \! +\!  1)\delta (t - t') \! + \! i \delta '(t - t') \right] ,
\label{E16}
\end{equation}
where ${{\bar n}_m} = {(\exp (\hbar {\omega _m}/{k_B}T) - 1)^{ - 1}}$ is the mean thermal phonon number and $\delta '(t - t')$ is the time derivative of the Dirac delta. The term proportional to the derivative of the Dirac delta is the antisymmetric part of the correlation function, associated with the commutator of $\eta(t)$ \cite{vitali noise membrane}, but it does not contribute to the subsequent expressions where we have always calculated \emph{symmetrized} correlation functions.

We define the optical and atomic quadrature operators ${{\hat X}_a} = ({{\hat a}^\dag } + \hat a)/\sqrt 2$, $\hat P_a = i({{\hat a}^\dag } - \hat a)/\sqrt 2$, ${{\hat X}_d } = ({\hat d } + {\hat d }^{\dagger})/\sqrt 2$, $ \hat P_d = i({\hat d }^{\dagger}- {\hat d })/\sqrt 2$ and their corresponding noise operators $\hat X_a^{in} = (\hat a^{in,\dag}  + {{\hat a}^{in}})/\sqrt 2$, $\hat P_a^{in} = i(\hat a^{in,\dag}  - {{\hat a}^{in}})/\sqrt 2$, $\hat X_d^{in} = (\hat d^{in,\dag} + \hat d^{in})/\sqrt 2$ and $\hat P_d^{in} = i(\hat d^{in,\dag } - \hat d^{in})/\sqrt 2$. Moreover we adopt dimensionless MO position and momentum operators $\hat {X} = \hat{x}/ \sqrt{2}x_{ZPF}$ and $\hat {P} = \hat{p}/ \sqrt{2}p_{ZPF}$, so that $[\hat {X},\hat {P}]=i$.
We then consider the usual regime where the cavity field and the atoms are strongly driven and the weak coupling optomechanical limit, so that we can linearize the dynamics of the quantum fluctuations around the semiclassical steady state. After straightforward calculations, the linearized quantum Langevin equations for the quadratures' fluctuations are obtained as
\begin{subequations} \label{fluctuation}
\begin{eqnarray}
&& \delta \dot{\hat X }= {\omega _m}\delta \hat P ,\\
&& \delta {{\dot {\hat X}}_d } =  - {\omega _m }\delta {{\hat P}_d } - \frac{\Gamma }{2}\delta {{\hat X}_d } + \sqrt \Gamma  \hat X_d ^{in} ,\\
&& \delta {{\dot {\hat X}}_a} = {\Delta _c}\delta {{\hat P}_a} - \frac{\kappa }{2}\delta {{\hat X}_a} + \sqrt \kappa  \hat X_a^{in} , \\
&& \delta \dot {\hat P} =\!  - {\omega _m}\delta \hat X \! - \! {\gamma _m}\delta \hat P \! - g\delta {{\hat X}_a} \! + \! \sqrt {{\gamma _m}} (\hat f \! +\! {F_{ext}}) ,\\
&& \delta {{\dot {\hat P}_a}} =\! - \! {\Delta _c}\delta {{\hat X}_a} \!  - \! g\delta \hat X \! - \! G\delta {{\hat X}_d } \! - \! \frac{\kappa }{2}\delta {{\hat P}_a} + \sqrt \kappa  \hat P_a^{in}\! ,\\
&& \delta {{\dot{ \hat P}}_d } = {\omega _m }\delta {{\hat X}_d } - G\delta {{\hat X}_a} - \frac{\Gamma }{2}\delta {{\hat P}_d } + \sqrt \Gamma  \hat P_d ^{in} ,
\end{eqnarray}
\end{subequations}
where the effective linearized optomechanical coupling constant is $g =  2 g_0 \alpha_s$, ${\Delta _c}=\Delta_{c0}-g_0^2 |\alpha_s|^2/\omega_m$ is the effective cavity detuning, and $\alpha_s$ is the intracavity field amplitude, solution of the nonlinear algebraic equation $(\kappa/2+i\Delta_c)\alpha_s = E_L -i G^2 \omega_m {\rm Re}\alpha_s/(\Gamma^2/4 + \omega_m^2)$, which is always possible to take as a real number by an appropriate redefinition of phases. Finally we have rescaled the thermal and external force by defining $f(t)= \eta(t)/\sqrt {\hbar m{\omega _m}{\gamma _m}}$ and $F_{\rm ext}=\tilde{F}_{\rm ext}(t)/\sqrt {\hbar m{\omega _m}{\gamma _m}}$.
These equations are analogous to those describing the CQNC scheme proposed in \cite{PRX} and then adapted to the case when the NMO is realized by a blue detuned cavity mode in Ref. \cite{maximilian}, and by an inverted atomic ensemble in Ref. \cite{meystre}. Compared to the latter paper, the cavity mode tunnel splitting $2J$ is replaced by the effective cavity mode detuning $\Delta_c$.

As suggested by the successful example of the injection of squeezing in the LIGO detector \cite{LIGO} and more recently in an electro-mechanical system \cite{Clark}, we now show that the force detection sensitivity of the present scheme can be further improved and can surpass the SQL when the cavity is driven by a squeezed vacuum field, with a spectrum centered at the cavity resonance frequency $\omega_{sq}=\omega_c$.
%in parallel to the classical coherent input field.

The squeezed field driving is provided by the finite bandwidth output of an optical parametric oscillator (OPO), shined on the input of our cavity system, implying that the cavity mode is subject to a non-Markovian squeezed vacuum noise, with two-time correlation functions given by \cite{55}
\begin{subequations}\label{squeezing noise}
\begin{eqnarray}
&& \left\langle {{{\hat a}^{in}}(t){{\hat a}^{in}}(t')} \right\rangle \!  =\! \frac{M}{2}\! \frac{{{b_x}{b_y}}}{{b_x^2 + b_y^2}}\left( {{b_y}{e^{ - {b_x}\tau }}\! +\! {b_x}{e^{ - {b_y}\tau }}} \right)\! , \\
&& \left\langle {\hat a^{in,\dag}(t){{\hat a}^{in}}(t')} \right\rangle \! = \! \frac{N}{2}\! \frac{{{b_x}{b_y}}}{{b_y^2 - b_x^2}}\!\left( {{b_y}{e^{ \!-{b_x}\tau }}\! -\! {b_x}{e^{\! - {b_y}\tau }}} \right)\! ,
\end{eqnarray}
\end{subequations}
where $\tau  = \left| {t - t'} \right|$, while $b_x$ and $b_y$ define the bandwidth properties of the OPO driven below threshold \cite{56} for the generation of squeezed light. The squeezing bandwidths and the parameters $M$ and $N$ are related to the effective second-order nonlinearity $\varepsilon$  and the cavity decay rate $\gamma$ of the OPO by ${b_{x}} = \gamma /2 - \left| \varepsilon  \right|$, ${b_{y}} = \gamma /2+ \left| \varepsilon  \right|$  and $M = (\varepsilon \gamma /2)(1/{b_x^2} + 1/{b_y^2})$, $N = (\left| \varepsilon  \right|\gamma /2)(1/{b_x^2} - 1/{b_y^2})$. It is clear that $N \ge 0$ and the stability of the OPO requires ${b_x} \ge 0$. The chosen parametrization satisfies the well known condition ${\left| M \right|^2} \le N(N + 1)$ for squeezed noise. In the case of pure squeezing, there are only two independent parameters, one can parametrize $M = (1/2)\sinh (2r) \exp({  i\phi })$ and $N = {\sinh ^2}r$, with $r$ and $\phi$ being, respectively, the strength and the phase of squeezing, so that ${\left| M \right|^2} = N(N + 1)$ and ${b_y} = {b_x}\sqrt {2(N + \left| M \right| + 1)}$.

In the white noise limit, i.e., when ${b_{x,y}} \to \infty$, while keeping $M$ and $N$ constant, the correlation functions can be written in Markovian form, i.e., $\left\langle {{{\hat a}_{in}}(t){{\hat a}_{in}}(t')} \right\rangle \! =\! M\delta (t - t')$ and $\left\langle {\hat a^{in,\dag} (t){{\hat a}^{in}}(t')} \right\rangle \! =\! N\delta (t - t')$. We will restrict to this white noise limit from now on, which is justified whenever the two bandwidths ${b_{x,y}}$ are larger than the mechanical frequency $\omega_m$ and the cavity line-width $\kappa$.

In the next section, we will study how CQNC eliminating the effect of backaction noise and squeezed-vacuum injection can jointly act in order to improve significantly the detection of a weak force acting on the MO.

%%%%%%%%%%%%%%%%%%%%%%%%%%%%%%%%%%%%%%%%%%%%%%%%%
\section{\label{sec4}Force sensing and CQNC}

An external force acting on the MO shifts its position, which in turn is responsible for a change of the effective length of the cavity and therefore of the phase of the optical cavity output. As a consequence, the signal associated to the force can be extracted by measuring the optical output phase quadrature, $\hat P_a^{out}$, with heterodyne or homodyne detection. The expression for the output filed can be obtained from the standard input-output relation \cite{GZ,input-output}, i.e., $\hat a^{out}= \sqrt{\kappa} \delta \hat a - \hat a^{in}$, so that the output quadrature is given by
\begin{equation}
\hat P_a^{out} = \sqrt \kappa  \delta {{\hat P}_a} - \hat P_a^{in} ,
\label{input-output}
\end{equation}
and solving Eqs. (\ref{fluctuation}) for $\delta {{\hat P}_a}$. Typically stationary spectral measurements of forces are carried out and therefore we are interested in the solution for $P_a^{out}$ in the frequency domain. After straightforward calculations, we get
\begin{eqnarray}
&&\hat P_a^{out} = \sqrt \kappa  {{\chi '}_a}\left\{ { - g{\chi _m}\sqrt {{\gamma _m}} \left( {\hat f + {F_{\rm ext}}} \right)} \right.\nonumber \\
&&\qquad \quad+\sqrt \kappa  \left[ {\left(1 - \frac{1}{{{{\chi '}_a}\kappa }}\right)\hat P_a^{in} - {\Delta _c}{\chi _a}\hat X_a^{in}} \right] \nonumber \\
&&\qquad\quad - G{\chi_d }\sqrt \Gamma  \left[ {\hat P_d ^{in} - \hat X_d^{in}\left(\frac{{\Gamma /2 + i\omega }}{{{\omega _m }}}\right)} \right] \nonumber \\
&&\qquad\quad +\left. \sqrt \kappa  {\chi _a}\left({g^2}{\chi _m} + {G^2}{\chi_{d }} \right) \hat X_a^{in} \right\} ,
\label{Pout}
\end{eqnarray}
where we have defined the susceptibilities of the cavity field, the MO, and of the atomic ensemble, respectively as
\begin{eqnarray}
&&{\chi _a}(\omega ) = \frac{1}{{\kappa /2 + i\omega }} ,\nonumber \\
&&{\chi _m}(\omega ) = \frac{{{\omega _m}}}{{\left( {\omega _m^2 - {\omega ^2}} \right) + i\omega {\gamma _m}}} ,\nonumber \\
&&{\chi_d }(\omega ) = \frac{{ - {\omega _m }}}{{\left( {\omega _m ^2 - {\omega ^2} + {\Gamma ^2}/4} \right) + i\omega \Gamma }} ,
\label{susceptibilities}
\end{eqnarray}
and we have introduced the modified cavity mode susceptibility
\begin{equation}
\frac{1}{{{{\chi '}_a}}} = \frac{1}{{{\chi _a}}} - {\chi _a}{\Delta _c}\left( {{g^2}{\chi _m} + {G^2}{\chi_d } - {\Delta _c}} \right) .
\label{effectivesusceptibility}
\end{equation}
Eq. (\ref{Pout}) is the experimental signal which, after calibration, is used for estimating the external force $F_{\rm ext}$. Appropriately rescaling Eq. (\ref{Pout}) we can rewrite
\begin{equation}
{F_{\rm ext}^{est}} \equiv \frac{{ - 1}}{{g{{\chi '}_a}{\chi _m}\sqrt {\kappa {\gamma _m}} }}\hat P_a^{out} \equiv {F_{\rm ext}} + {\hat F_{N}},
\label{Force}
\end{equation}
where the added force noise is defined as
\begin{eqnarray}
&&{{\hat F}_{N}} = \hat f - \sqrt {\frac{\kappa }{{{\gamma _m}}}} \frac{1}{{g{\chi _m}}}\left[ {\left(1 - \frac{1}{{{{\chi '}_a}\kappa }}\right)\hat P_a^{in} - {\Delta _c}{\chi _a}\hat X_a^{in}} \right] \nonumber \\
&&\qquad\qquad\qquad + \frac{{G{\chi_d }}}{{g{\chi _m}}}\sqrt {\frac{{{\Gamma }}}{{{\gamma _m}}}} \left[ {\hat P_d ^{in} - \frac{{\Gamma /2 + i\omega }}{{{\omega_m }}} \hat X_d^{in}} \right] \nonumber \\
&&\qquad\qquad\qquad - \frac{{{g^2}{\chi _m} + {G^2}{\chi_d }}}{{g{\chi _m}}}\sqrt {\frac{\kappa }{{{\gamma _m}}}} {\chi _a}\hat X_a^{in}.
\label{added force}
\end{eqnarray}
Eq. (\ref{added force}) shows that in the present scheme for force detection we have four different contributions to the force noise spectrum. The first term corresponds to the thermal noise of the MO, the second term corresponds to the shot noise associated with the output optical field, which is the one eventually modified by the squeezed input field. The third term is the contribution of the atomic noise due to its interaction with the cavity mode, while the last term describes the backaction noise due to the coupling of the intracavity radiation pressure with the MO and with the atomic ensemble.

\subsection{CQNC conditions}
The CQNC effect amounts to the perfect backaction cancellation at all frequencies, obtaining in this way significantly lower noise in force detection. From the last term in Eq. (\ref{added force}), it is evident that for $g=G$ and $\chi_m=-\chi_{d}$ the contributions of the backaction from the mechanics and from the atoms cancel each other for all frequencies. As shown in Fig. \ref{fig1}(b), they can be thought of as `noise' and `anti-noise' path contributions to the signal force $F_{\rm ext}$. Therefore an effective NMO, in this case realized by the inverted atomic ensemble, is necessary for realizing $\chi_m=-\chi_{d}$. More in detail, CQNC is realized whenever:
\begin{itemize}
\item[(i)]
the coupling constant of the optical field with the MO and with the atomic ensemble are perfectly matched, $g=G$, which is achievable by adjusting the intensity of the fields driving the cavity and the atoms;
\item[(ii)]
the atomic dephasing rate between the two lower atomic levels $\Gamma$ must be perfectly matched with the mechanical dissipation rate $\gamma_m$ (we have assumed the atomic Zeeman splitting perfectly matched with the MO frequency $\omega_m$ from the beginning);
\item[(iii)]
the MO has a high mechanical quality factor, or equivalently, $\Gamma  \ll {\omega _m}$ so that the term ${\Gamma ^2}/4$ can be neglected in the denominator of $\chi_{d} $ (see Eq. (\ref{susceptibilities})).
\end{itemize}
Mechanical damping rates of high quality factor MO are quite small, not larger than 1 kHz. As already pointed out in Sec. III of Ref. \cite{meystre}, the matching of the two decay rates is easier in the case of atoms because ground state dephasing rates can also be quite small \cite{heinze,dudin}. On the contrary, matching the dissipative rates in the case when the NMO is a second cavity mode, as in the fully optical model of Ref. \cite{maximilian}, is more difficult because it requires having a cavity mode with an extremely small bandwidth which can be obtained only assuming large finesse and long cavities.

Note that under CQNC conditions the effective susceptibility of Eq. (\ref{effectivesusceptibility}) becomes ${\chi '}_a^{CQNC} = (1/{\chi _a} + {\chi _a}\Delta _{c}^2)^{ - 1}$.
It is clear that under the CQNC conditions the last term in the noise force of Eq. (\ref{added force}) is identically zero and we can rewrite
\begin{eqnarray}
&&{{\hat F}_{N}} = \hat f - \sqrt {\frac{\kappa }{{{\gamma _m}}}} \frac{1}{{g{\chi _m}}}\left[ {\left(1 - \frac{1}{{{{\chi '}_a}\kappa }}\right)\hat P_a^{in} - {\Delta _c}{\chi _a}\hat X_a^{in}} \right] \nonumber\\
 &&\qquad\quad- \left[ {\hat P_d^{in} - \frac{{\Gamma /2 + i\omega }}{{{\omega _m }}} \hat X_d^{in}} \right] .
\label{FadeCQNC}
\end{eqnarray}

In order to quantify the sensitivity of the force measurement, we consider the spectral density of added noise which is defined by \cite{maximilian}
\begin{equation}
{S_F}(\omega )\delta (\omega - \omega ') = \frac{1}{2}\left( {\left\langle {\hat F_{N}(\omega )\hat F_{N}(-\omega ')} \right\rangle  + c.c} \right).
\label{spectrum}
\end{equation}
Under perfect CQNC conditions one gets the force noise spectrum in the presence of squeezed-vacuum injection which, in the experimentally relevant case $\kappa \gg \omega$, reads (see Appendix A for the explicit derivation)
\begin{eqnarray}
&&{S_F}(\omega ) = \frac{{{k_B}T}}{{\hbar {\omega _m}}} + \frac{1}{2}\left( {1 + \frac{{{\omega ^2} + \gamma _m^2/4}}{{\omega _m^2}}} \right) \nonumber\\
 &&\qquad \quad \quad \! + \frac{\kappa }{{{g^2}{\gamma _m}}}\!\frac{1}{{{{\left| {{\chi _m}}\! \right|}^2}}}\! \left[ \frac{1}{2}\! \left(\! \frac{1}{2}\! +\! \frac{2 \Delta_c^2}{\kappa^2 }\right)^2 \! \! \!+ \! \Sigma (M,N,\Delta _c/\kappa) \right]\! , \nonumber \\
\label{final spectrum}
\end{eqnarray}
where
\begin{eqnarray}
&&\Sigma (M,N,\Delta _c/\kappa ) = N{{\left(\frac{1}{2} +\!  {{\frac{{{2 \Delta_c^2}}}{\kappa^2 }}}\right)}^2} \! + 2\frac{{{\Delta _c}}}{\kappa }{\mathop{\rm Im}\nolimits} M\left(4\frac{{\Delta _c^2}}{{{\kappa ^2}}} - 1\right) \nonumber \\
 &&\qquad\qquad\qquad \qquad + {\mathop{\rm Re}\nolimits} M\left[ \frac{8 \Delta _c^2}{\kappa ^2} - \left(\frac{1}{2} + \frac{2 \Delta_c^2}{\kappa^2 }\right)^2 \right]
\label{squeezingterm}
\end{eqnarray}
is the contribution of the injected squeezing to the optomechanical shot noise.
Eq. (\ref{final spectrum}) shows that when CQNC is realized, the noise spectrum consists of three contributions: the first term denotes the thermal Brownian noise of the MO, the second term describes the atomic noise, and the last one represents the optomechanical shot noise modified by squeezed-vacuum injection.

We recall that with the chosen units, the noise spectral density is dimensionless and in order to convert it to ${N^2}H{z^{ - 1}}$ units we have to multiply by the scale factor $\hbar m{\omega _m}{\gamma _m}$.
This noise spectrum has to be compared with the noise spectrum of a standard optomechanical setup formed by a single cavity coupled to a MO at resonance frequency ($\Delta_c=0$) \cite{5,7},
\begin{equation}
{S_F^{st}}(\omega ) = \frac{{{k_B}T}}{{\hbar {\omega _m}}} + \frac{1}{2}\left[ {\frac{1}{4}\frac{\kappa }{{{g^2}{\gamma _m}}}\frac{1}{{{{\left| {{\chi _m}} \right|}^2}}} + 4{g^2}\frac{1}{{\kappa {\gamma _m}}}} \right].
\label{sopt}
\end{equation}
As it is well known, the standard quantum limit for stationary force detection comes from the minimization of the noise spectrum at a given frequency over the driving power, i.e., over the linearized coupling squared ${{g^2}}$, yielding
\begin{equation}
{S_F^{st}}(\omega ) \ge {S_{SQL}} = \frac{1}{{{\gamma _m}\left| {{\chi _m}(\omega )} \right|}}.
\label{SQL}
\end{equation}
In the present case, the complete cancellation of the backaction noise term proportional to $g^2$ has the consequence that force detection is limited only by shot noise and that therefore the optimal performance is achieved at very large power. In this limit force detection is limited only by the
the additional shot-noise-type term that is independent of the measurement strength ${{g^2}}$ corresponding to atomic noise (see Eq. (\ref{final spectrum})), and which is the price to pay for the realization of CQNC,
\begin{equation}
{S_{CQNC}} = \frac{1}{2}\left( {1 + \frac{{{\omega ^2} + \gamma _m^2/4}}{{\omega _m^2}}} \right),
\label{sCQNC}
\end{equation}
(here we neglect thermal noise and other technical noise sources which are avoidable in principle). As already discussed in Ref. \cite{meystre}, in the limit of sufficiently large driving powers when shot noise (and also thermal noise) is negligible, CQNC has the advantage of significantly increasing the bandwidth of quantum-limited detection of forces, well out of the mechanical resonance.
This analysis can be applied also for the present scheme employing a single cavity mode, and it is valid also in the presence of injected squeezing, which modifies and can further suppress the shot noise contribution. This is relevant because it implies that one can achieve the CQNC limit of Eq. (\ref{sCQNC}), by making the shot noise contribution negligible, much easily, already at significantly lower driving powers. In this respect one profits from the ability of injected squeezing to achieve the minimum noise at lower power values, as first pointed out by Caves \cite{caves80}.

Let us now see in more detail the effect of the injected squeezing by optimizing the parameters under perfect CQNC conditions. To be more specific, the injected squeezed light has to suppress as much as possible the shot noise contribution to the detected force spectrum, and therefore we have to minimize the function within the square brackets of Eq. (\ref{final spectrum}), over the squeezing parameters $N$, $M$ and the detuning $\Delta_c$. Defining $y=\Delta_c/\kappa$ the normalized detuning, one can rewrite this function as
\begin{eqnarray}
&& h(M,N,y) = \left(N + \frac{1}{2}\right){\left(\frac{1}{2} + 2{y^2}\right)^2} \nonumber \\
&&\qquad \qquad\qquad - |M|\left[a(y) \sin \phi + b(y) \cos\phi\right],
\end{eqnarray}
where $M=|M|e^{i\phi}$, and we have introduced the detuning-dependent functions
\begin{eqnarray}
a(y)&=&2y(1-4{y^2}) \nonumber \\
b(y)&=&\left(\frac{1}{2} + 2y^2\right)^2 -8y^2 .
\label{Fsimple}
\end{eqnarray}
$h(M,N,y)$ can be further rewritten as
\begin{eqnarray}
&& h(M,N,y) = \left(N + \frac{1}{2}\right){\left(\frac{1}{2} + 2{y^2}\right)^2} \nonumber \\
&&\qquad \qquad\qquad -|M|\sqrt{a(y)^2+b(y)^2}\cos\left[\phi-\phi_{\rm opt}(y)\right], \nonumber  \\
\end{eqnarray}
where $\tan \phi_{\rm opt}(y)=a(y)/b(y)$ and it is straightforward to verify that $\sqrt{a(y)^2+b(y)^2}=\left(1/2 + 2y^2\right)^2$. From this latter expression is evident that, for a given detuning $y$, and whatever value of $N$ and $|M|$, the optimal value of the squeezing phase minimizing the shot noise contribution is just $\phi=\phi_{\rm opt}(y)$, for which one gets
\begin{equation} \label{h_general}
h(|M|,N,y) = \left(N + \frac{1}{2}-|M|\right){\left(\frac{1}{2} + 2{y^2}\right)^2} .
\end{equation}
This latter expression can be easily further minimized by observing that its minimal value is obtained by assuming pure squeezed light $|M|=\sqrt{N(N+1)}$ and also taking zero detuning $y=0$, i.e., driving the cavity mode at resonance, so that for a given value of (pure) squeezing $N$, one has that
\begin{equation}\label{h_particular}
h_{\rm min}(N) = \frac{1}{4} \left[N+1/2-\sqrt{N(N+1)}\right],
\end{equation}
which tends to zero quickly for large squeezing $N$, i.e., $h_{\rm min}(N \gg 1) \to 1/(32N)$. As a consequence, the shot noise contribution can be rewritten after optimization over the squeezing and detuning as,
\begin{equation}\label{speopt}
{S_F}^{\rm shot, opt}(\omega )=\frac{\kappa }{4 g^2\gamma _m \left| \chi _m \right|^2} \left[N+1/2-\sqrt{N(N+1)}\right] .
\end{equation}
We notice that the optimal value of the detuning, $\Delta_c=0$ can be taken only in the present model with a single cavity mode and not in the dual-cavity model of Ref.~\cite{meystre} where the parameter $\Delta_c$ is replaced by the coupling rate between the two cavities $2J$, which cannot be reduced to zero. This is an important advantage of the single cavity mode case considered here.
Eq.~(\ref{speopt}) shows that injected squeezing greatly facilitates achieving the ultimate limit provided by CQNC of Eq. (\ref{sCQNC}) because in the optimal case and at large squeezing $N$, the shot noise term is suppressed by a factor $1/(4N)$ with respect to the case without injected squeezing (compare Eq. (\ref{h_general}) in the case $M=N=y=0$ with Eq. (\ref{h_particular}) in the case when $N \gg 1$). This is of great practical utility because it means that one needs a much smaller value of $g$, and therefore much less optical driving power in order to reach the same suppression of the shot noise contribution.

\begin{figure}
\begin{center}
\includegraphics[width=8.5cm]{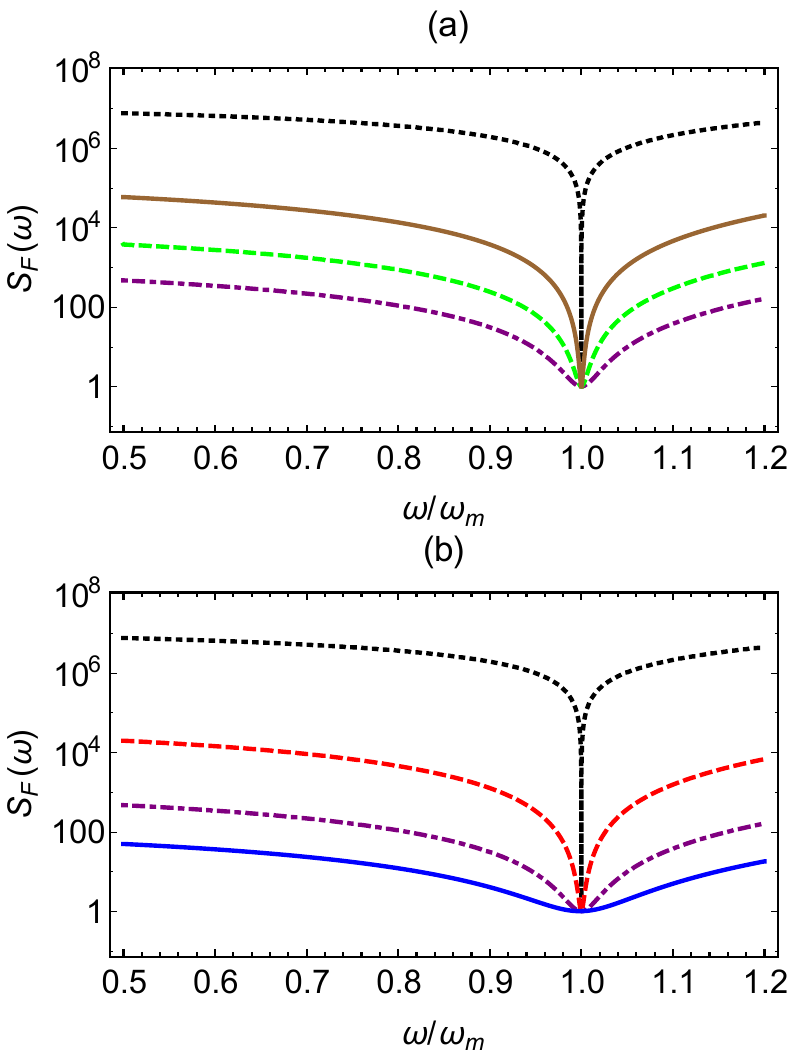}
\end{center}
\caption{(Color online) Force noise spectral density versus $\omega/\omega_m$ in the presence of perfect CQNC, with an optimized squeezed injected light with phase $\phi= \phi_{\rm opt}(0) = 0$ and $|M|=\sqrt{N(N+1)}$. (a) refers to the case with fixed squeezing $N=10$ and different detunings: $\Delta_c/\kappa=0$ (dot-dashed purple line), $\Delta_c/\kappa=1/2$ (dashed green line), $\Delta_c/\kappa=1$ (full brown line). (b) refers to the optimal case $\Delta_c/\kappa=0$ and different values of the squeezing parameter, $N=0$ (dashed, red line), $N=10$ (dot-dashed, purple line), and $N=100$ (full, blue line). The dotted black line corresponds to the SQL. The other parameters are ${\omega _m}/2\pi  = 300$ KHz, ${\gamma _m}/2\pi  = 30 $ mHz, ${g_0}/2\pi  = 300$ Hz, ${\lambda _L} \simeq 780$ nm, $P_L=24 \mu $W and $\kappa/2\pi=1$ MHz.}
\label{fig2}
\end{figure}

Let us now illustrate how the combination of backaction cancellation by the atomic ensemble under CQNC and of the squeezing injected in the cavity may significantly improve force detection sensitivity. We consider an experimentally feasible scheme based on a membrane-in-the-middle setup \cite{thompson}, coupled to an ultracold atomic gas confined in the cavity and in a magnetic field, like the one demonstrated in Ref. \cite{dudin} for light storage. A system of this kind has not been demonstrated yet, but the coupling of an atomic ensemble with a membrane has been already demonstrated in Refs. \cite{40,42}. We assume typical mechanical parameter values for SiN membranes, ${\omega _m}/2\pi  = 300$ KHz, ${\gamma _m}/2\pi  = 30 $ mHz, ${g_0}/2\pi  = 300$ Hz, ${\lambda _L} \simeq 780$ nm, $P_L=24 \mu $W and $\kappa/2\pi=1$ MHz (see also the caption of Fig. 2). The ground state sub-levels of the ultracold atomic gas of Ref. \cite{dudin} could be prepared in order to satisfy the CQNC condition, i.e., the Zeeman splitting tuned in order that the effective atomic transition rate coincides with $\omega_m$, the driving of the laser fields adjusted so that the two linearized couplings with the cavity mode, $G$ and $g$, coincide. Matching the dephasing rate $\Gamma$ with the damping rate $\gamma_m$ is less straightforward but one can decrease and partially tune the atomic dephasing rate using the magic-value magnetic field technique and applying dynamical decoupling pulse sequences, as demonstrated in Ref. \cite{dudin}.

In Fig. (\ref{fig2}), the force noise spectral density ${S_F}(\omega )$ optimized over the squeezing parameters, i.e., $|M| = \sqrt{N(N+1)}$, $\phi =0$, is plotted versus frequency. In Fig. \ref{fig2}(a) we fix the squeezing parameter $N=10$ and consider different values of the detuning: as shown above, force noise is minimum at the optimal case of resonant cavity driving $\Delta_c = 0$. This plot clearly shows the advantage of the present single cavity scheme compared to the double cavity setup of Ref. \cite{meystre}, where the role of $\Delta_c$ is played by the mode splitting $2J$ associated with the optical coupling $J$ between the cavity that cannot be put to zero. In Fig. \ref{fig2}(b) we fix the detuning at this optimal zero value, and we consider different values of the injected squeezing parameter $N$. At resonance ($\omega=\omega_m$), CQNC and injected squeezing does not improve with respect to the SQL spectrum (dotted black line), but force noise suppression is remarkable in a broadband around the resonance peak, and becomes more relevant for increasing injected squeezing $N$. Notice that injected squeezing allows a further reduction of the off-resonance ($\omega \neq \omega_m$) force noise with respect to what can be achieved with CQNC alone (see in Fig. \ref{fig2}(b) the full blue line and the dot-dashed purple line compared to the dashed red line which refers to the case of no-injected squeezing, $N=0$.)

\begin{figure}
\begin{center}
\includegraphics[width=8.5cm]{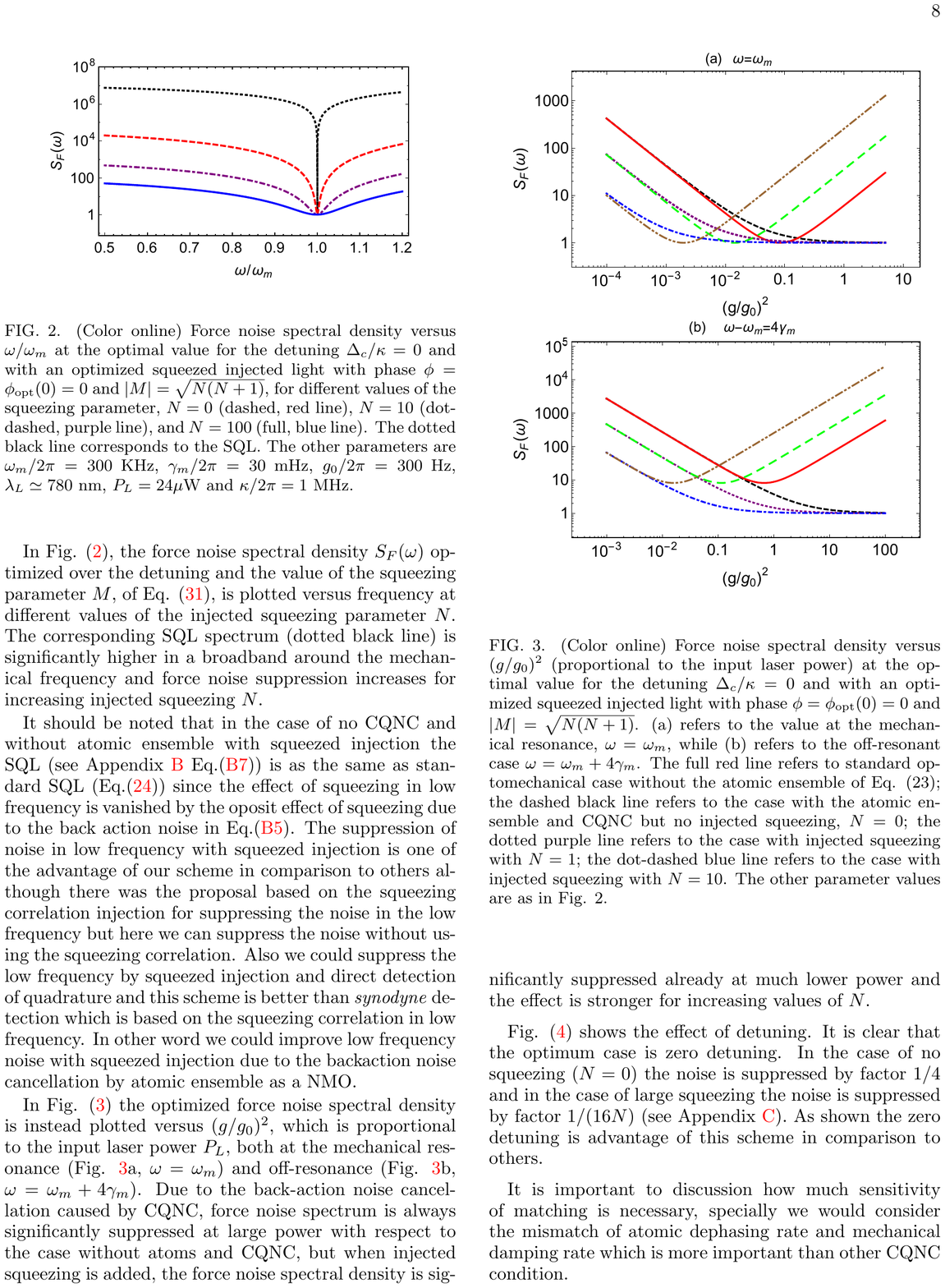}
\end{center}
\caption{(Color online) Force noise spectral density versus ${(g/{g_0})^2}$ (proportional to the input laser power) at the optimal value for the detuning $\Delta_c/\kappa=0$ and with an optimized squeezed injected light with phase $\phi= \phi_{\rm opt}(0) = 0$ and $|M|=\sqrt{N(N+1)}$. (a) refers to the value at the mechanical resonance, $\omega=\omega_m$, while (b) refers to the off-resonant case $\omega=\omega_m+4\gamma_m$. In both subfigures we compare the force noise spectrum with perfect CQNC and for a given (optimized) squeezing $N$ with the corresponding spectrum with the same injected squeezing but without atomic ensemble and CQNC. The full red line refers to standard optomechanical case $N=0$ without the atomic ensemble of Eq. (\ref{sopt}), while the dashed black line refers again to $N=0$ \emph{with} the atomic ensemble and perfect CQNC. The case with injected squeezing with $N=1$ corresponds to the long-dashed green line (without atoms and CQNC, see also Eq. (\ref{s_optomechanic standard with squeezing})), and to the dotted purple line (with atoms and CQNC). Finally the case with injected squeezing with $N=10$ corresponds to the double-dot dashed brown line (no atoms and CQNC), and to the dot-dashed blue line (with atoms and perfect CQNC). The other parameter values are as in Fig. (\ref{fig2}).}
\label{fig3}
\end{figure}

In Fig. (\ref{fig3}) the force noise spectral density is instead plotted versus $(g/g_0)^2$, which is proportional to the input laser power $P_L$, both at the mechanical resonance (Fig. \ref{fig3}(a), $\omega=\omega_m$) and off-resonance (Fig. \ref{fig3}(b), $\omega=\omega_m+4\gamma_m$). In both subfigures we compare the force noise spectrum with perfect CQNC and for a given optimized squeezing $N$, with the corresponding spectrum with the same injected squeezing but without atomic ensemble and CQNC, for three different values of $N$, $N=0, 1, 10$ (see also Appendix B where we evaluate the general expressions of the force noise spectrum without imposing the CQNC condition).
Back-action noise cancellation manifests itself with a significant noise suppression at large power, where minimum force noise is achieved. Without atoms and CQNC, force noise diverges at large power due to backaction, and one has the usual situation where minimum force noise is achieved at the SQL, at a given optimal power. In both cases, either with or without CQNC, injected squeezing with $\phi= 0$ and $|M|=\sqrt{N(N+1)}$ is not able to improve force sensitivity and to lower the noise at resonance (see Fig. \ref{fig3}(a)), i.e., the SQL value remains unchanged, but one has the advantage that for increasing $N$, the SQL is reached at decreasing values of input powers \cite{caves80}. As already suggested by Fig. (\ref{fig2}), instead one has a significant force noise suppression off-resonance and at large powers due to backaction cancellation (Fig. \ref{fig3}(b)).

\subsection{The case of imperfect CQNC conditions}

Backaction cancellation requires the perfect matching of atomic and mechanical parameters. As we discussed above, one can tune the effective atomic transition rate by tuning the magnetic field, and make it identical to the mechanical resonance frequency $\omega_m$. Here we still assume such a perfect frequency matching which, even though not completely trivial, can always be achieved due to the high tunability of Zeeman splitting. As we discussed in the previous subsection, one can also make the two couplings with the cavity mode $G$ and $g$ identical, by adjusting the cavity and atomic driving, and finally even the two decay rates, $\Gamma$ and $\gamma_m$. However, both coupling rates matching and decay rate matching are less straightforward, and therefore it is important to investigate the robustness of the CQNC scheme with respect to imperfect matching of these two latter parameters.

\begin{figure}
\begin{center}
\includegraphics[width=8.5cm]{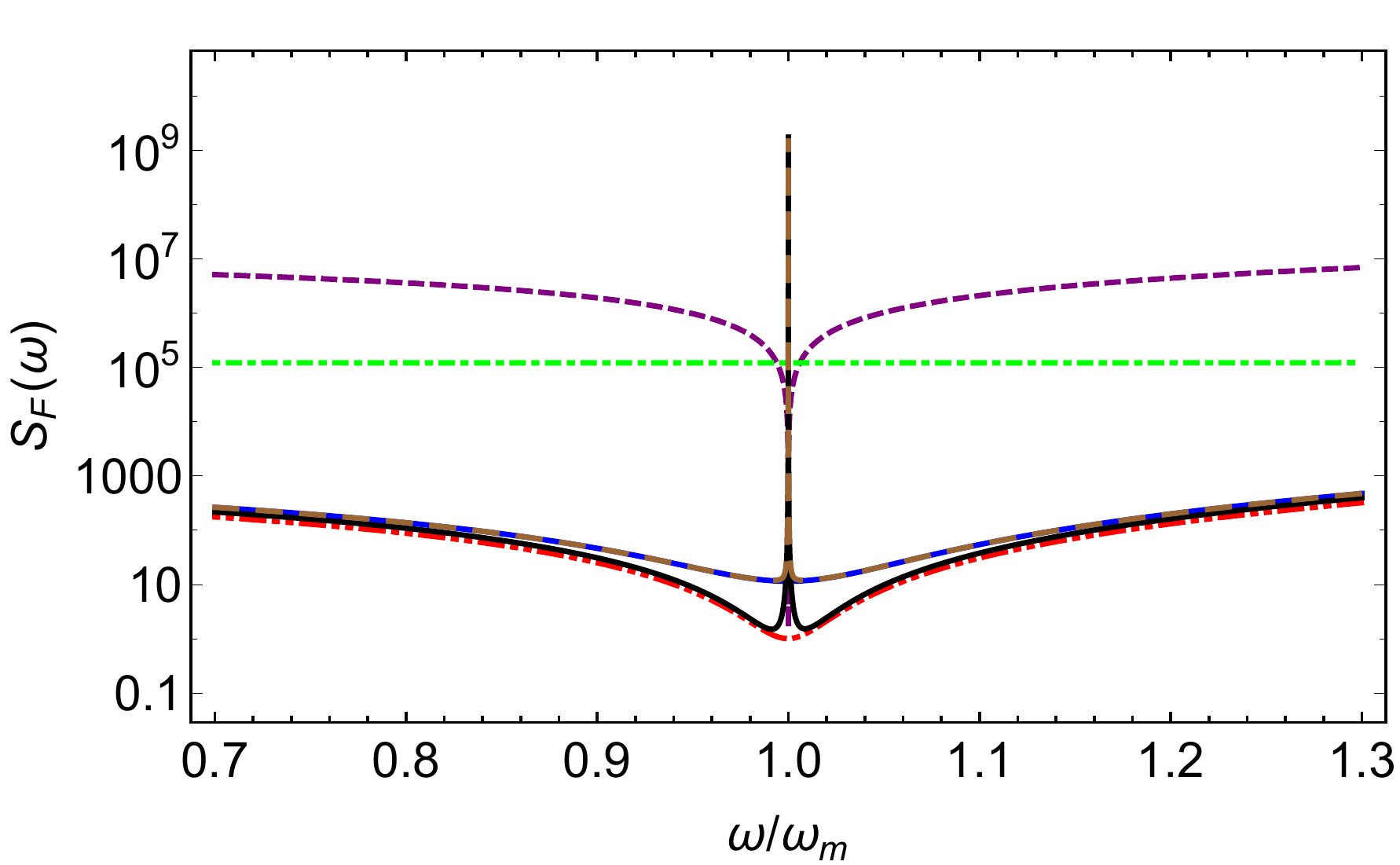}
\end{center}
\caption{(Color online)  Force noise spectral density versus $\omega/\omega_m$ at the optimal value for the detuning $\Delta_c/\kappa=0$, with an optimized squeezed injected light with phase $\phi= \phi_{\rm opt}(0) = 0$, $|M|=\sqrt{N(N+1)}$, and $N=10$. We consider different coupling and decay rate mismatches. The dashed purple line and double-dot dashed red line, respectively refer to the SQL and perfect CQNC. The other curves correspond to: $(G-g)/g=\pm 10^{-3}$, $\Gamma=\gamma_m$ (green, dot-dashed line); $(G-g)/g=\pm 10^{-5}$, $\Gamma=\gamma_m$ (blue, solid line); $(\Gamma-\gamma_m)/\gamma_m=\pm 0.5$, $g=G$ (black, solid line); $(G-g)/g=\pm 10^{-5}$, $(\Gamma-\gamma_m)/\gamma_m=\pm 0.5$ (brown, long-dashed line). The other parameter values are as in Fig. (\ref{fig2}).}
\label{fig4}
\end{figure}

We have restricted our analysis to the parameter regime  corresponding to the optimal case under perfect CQNC conditions, i.e., the resonant case  $\Delta_c=0$, with an optimized pure squeezing, $\phi= \phi_{\rm opt}(0) = 0$ and $|M|=\sqrt{N(N+1)}$. We have also fixed the squeezing value, $N= 10$, and considered again the parameter values of the previous subsection, but now considering the possibility of nonzero mismatch of the couplings and/or of the decay rates. We have used the expression of the spectrum of Eq. (\ref{S_zerodetuning}). We first consider in Fig. (\ref{fig4}) the effect of parameter mismatch on the force noise spectrum versus $\omega$. Fig. (\ref{fig4}) shows that CQNC is more sensitive to the coupling mismatch than to the decay rate mismatch. In fact, the spectrum is appreciably modified already when $(G-g)/g = 10^{-5}$, and force noise increases significantly and in a broadband around resonance already when $(G-g)/g = 10^{-3}$. This modification is quite independent from the value of the decay rate mismatch, $(\Gamma-\gamma_m)/\gamma_m$, whose effect moreover is always concentrated in a narrow band around resonance and for larger values, $(\Gamma-\gamma_m)/\gamma_m = 0.5$. There is a weak dependence upon the sign of the two mismatches, which however is typically very small and not visible in the plots.

\begin{figure}
\begin{center}
\includegraphics[width=8.5cm]{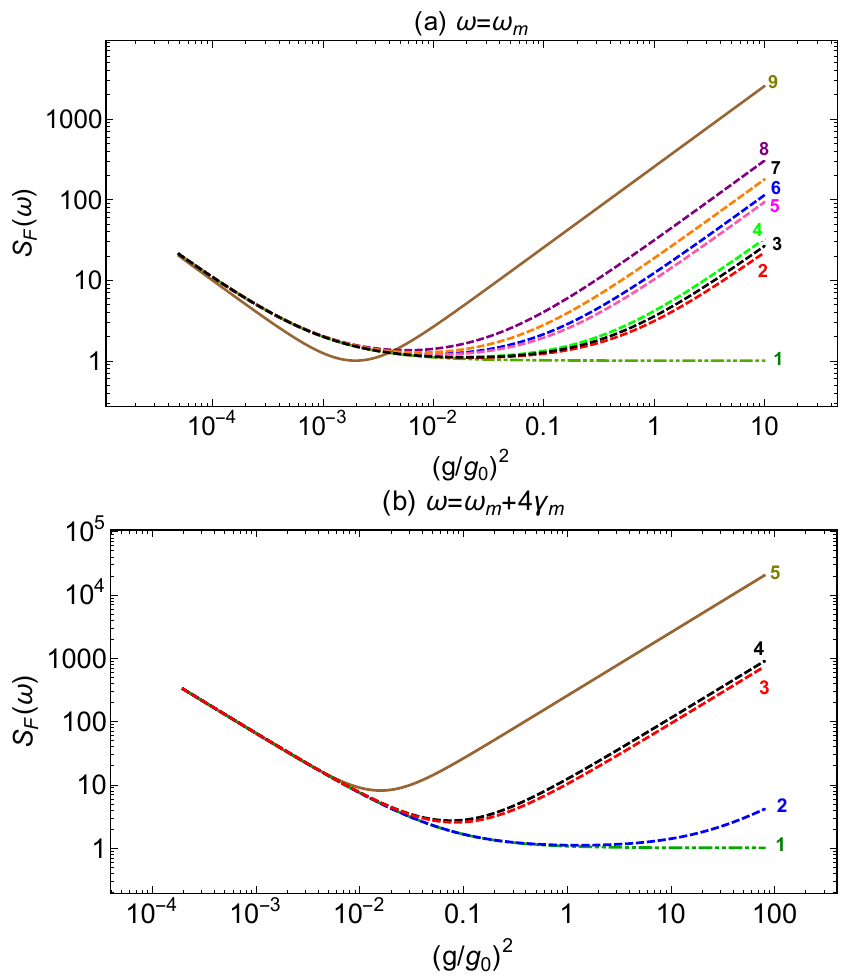}
\end{center}
\caption{(Color online) Force noise spectral density versus ${(g/{g_0})^2}$. (a) Resonant case ($\omega=\omega_m$). Curves 1 (double-dot-dashed dark green line) and 9 (brown solid line) refer to the CQNC and standard optomechanical spectrum (Eq. (\ref{s_optomechanic standard with squeezing})), respectively. The other curves correspond to:
$(\Gamma-\gamma_m)/\gamma_m=+0.1$, $G=g$ (curve 2, red dashed); $(\Gamma-\gamma_m)/\gamma_m=(G-g)/g=\pm 0.1$ (curve 3, black dashed);  $(\Gamma-\gamma_m)/\gamma_m=-0.1$, $g=G$ (curve 4, green dashed); $(G-g)/g=-0.1$, $\Gamma=\gamma_m$ (curve 5, pink dashed); $(G-g)/g=+0.1$, $\Gamma=\gamma_m$ (curve 6, blue dashed); $(G-g)/g=- (\Gamma-\gamma_m)/\gamma_m=-0.1$ (curve 7, orange dashed); and $(G-g)/g=- (\Gamma-\gamma_m)/\gamma_m=+0.1$ (curve 8, purple dashed).
Figure (b) refers to the off-resonant case ($\omega=\omega_m + 4\gamma_m$). Curves 1 (double-dot-dashed dark green line) and 5 (brown solid line) refer to the CQNC and standard optomechanical spectrum, respectively. The other curves correspond to: $(\Gamma-\gamma_m)/\gamma_m=\pm 0.1$ and $g=G$ (curve 2, blue dashed); $(G-g)/g=-0.1$ and $\Gamma=\gamma_m$, $(\Gamma-\gamma_m)/\gamma_m=(G-g)/g=-0.1$, and $(G-g)/g=- (\Gamma-\gamma_m)/\gamma_m=-0.1$ (curve 3, red dashed); $(G-g)/g=0.1$ and $\Gamma=\gamma_m$, $(\Gamma-\gamma_m)/\gamma_m=(G-g)/g=0.1$ and $(G-g)/g=- (\Gamma-\gamma_m)/\gamma_m=0.1$ (curve 4, black dashed). As in Fig. 4, all curves refer to $N=10$, $\Delta_c=0$, $\phi= \phi_{\rm opt}(0) = 0$ and $|M|=\sqrt{N(N+1)}$. The other parameter values are as in Fig. (\ref{fig2}).}
\label{fig5}
\end{figure}

In Fig. (\ref{fig5}) instead we fix the frequency and consider the dependence of the force noise spectrum versus $g^2$, i.e. versus the laser input power, either at resonance (Fig. \ref{fig5}(a)), and off-resonance (Fig. \ref{fig5}(b)), similarly to what we did under perfect CQNC in Fig. (\ref{fig3}). Due to the imperfect CQNC caused by parameter mismatch, at large power force noise spectrum increases again due to the uncancelled, residual backaction noise, and the increase at large power is larger for larger parameter mismatch. At resonance (Fig. \ref{fig5}(a)) both coupling mismatch and decay rate mismatch have an effect, and force noise increase is larger when both mismatches  are nonzero and opposite, due to the effect of the negative mass, yielding susceptibilities with opposite signs. As already shown in Fig. (\ref{fig4}), the effect of decay rate mismatch is instead hardly appreciable out of resonance, and noise increase is caused by the mismatch between the two couplings, regardless the value of the decay rate mismatch. The analysis of Figs. (\ref{fig4}) and (\ref{fig5}) allows us to conclude that CQNC is robust with respect to mismatch of the decay rates, up to $10\%$ mismatch, and especially off-resonance, where the advantage of backaction cancellation is more relevant. On the contrary, CQNC is very sensitive to the mismatch between the atomic and mechanical couplings with the cavity mode, which have to be controlled at $0.1 \%$ level or better. This means that in order to suppress backaction noise the intensity of the cavity and atomic driving have to be carefully controlled in order to adjust the two couplings.

\section{\label{sec5}summary and conclusion}
We have proposed a scheme for the realization of CQNC scheme for the high-sensitive detection of forces based on a single optomechanical cavity containing an atomic ensemble and driven by squeezed vacuum light.
The interaction of the atomic ensemble with the cavity mode leads to a destructive interference that perfectly cancels backaction noise of the MO, provided that atomic ensemble parameters are chosen so that it acts as a negative mass oscillator whose susceptibility perfectly cancels the mechanical one.
Perfect CQNC occurs when the optomechanical and the atom-field interaction coupling constants, the mechanical frequency and the effective atomic transition rate, and finally the dissipation rate of the mechanical resonator and the decoherence rate of the atomic ensemble, respectively coincide. The present scheme could be implemented by combining state-of-the-art membrane-in-the-middle setup \cite{thompson}, with ultracold atomic ensemble systems used for long-lived light storage \cite{dudin} and improves in various directions the dual cavity proposal of Ref. \cite{meystre}. The optical coupling rate between cavities $J$ in Ref. \cite{meystre} is replaced by the cavity mode detuning $\Delta_c$ in our scheme, and due to this fact, the present scheme reaches a stronger force noise suppression because such a suppression is optimal at resonance $\Delta_c = 0$, which can be set only in the present scheme. A further noise suppression is realized by injected squeezed vacuum in the cavity mode: in fact, shot noise is further suppressed for increased squeezing, and this occurs at much lower input laser power.
We have also analyzed in detail the effect of imperfect CQNC conditions, i.e., when the mechanical and atomic parameters are not perfectly matched, focusing on the case when the two couplings with the cavity modes and/or the decay rates are different. We have seen that backaction cancellation is robust withe respect to the decay rate mismatch and $10\%$ mismatch can be tolerated, especially off-resonance. Instead CQNC is very sensitive to the mismatch of the coupling rates, and one has to tune the two couplings, by adjusting the cavity and atomic driving, at the $0.1\% $ level at least.

%\begin{acknowledgements}
{\it Acknowledgments}
This work is supported by the European Commission through the Marie Curie ITN cQOM and FET-Open Project iQUOEMS. A. M. and F. B. wish to thank the Office of Graduate Studies of the University of Isfahan for their support.
%\end{acknowledgements}

\appendix

\section{\label{App.A} derivation of CQNC force noise spectrum}

Using the definitions provided in the main text, the force noise spectrum is explicitly written as
\begin{eqnarray}
&&\left\langle {{{\hat F}_{N}}(\omega ){{\hat F}_{N}}(-\omega ')} \right\rangle  = \nonumber \\
&&\qquad\left\langle {\hat f(\omega )\hat f( - \omega )} \right\rangle  + \frac{\kappa }{{{g^2}{\gamma _m}{\chi _m}(\omega ){\chi _m}( - \omega )}}\nonumber \\
&&\qquad\left[ {(1 - \frac{1}{{{{\chi '}_a}(\omega )\kappa }})(1 - \frac{1}{{{{\chi '}_a}( - \omega )\kappa }})} \right.\left\langle {\hat P_a^{in}(\omega )\hat P_a^{in}( - \omega )} \right\rangle \nonumber \\
&&\qquad - {\Delta _c}{\chi _a}( - \omega )(1 - \frac{1}{{{{\chi '}_a}(\omega )\kappa }})\left\langle {\hat P_a^{in}(\omega )\hat X_a^{in}( - \omega )} \right\rangle \nonumber \\
&&\qquad - {\Delta _c}{\chi _a}(\omega )(1 - \frac{1}{{{{\chi '}_a}( - \omega )\kappa }})\left\langle {\hat X_a^{in}(\omega )\hat P_a^{in}( - \omega )} \right\rangle \nonumber \\
&&\qquad\left. { + \Delta _c^2{\chi _a}(\omega ){\chi _a}( - \omega )\left\langle {\hat X_a^{in}(\omega )\hat X_a^{in}( - \omega )} \right\rangle } \right] \nonumber \\
 &&\qquad+ \left\langle {\hat P_d ^{in}(\omega )\hat P_d ^{in}( - \omega )} \right\rangle  - \frac{{\Gamma /2 - i\omega }}{{{\omega _m}}}\left\langle {\hat P_d ^{in}(\omega )\hat X_d ^{in}( - \omega )} \right\rangle \nonumber \\
&&\qquad- \frac{{\Gamma /2 + i\omega }}{{{\omega _m}}}\left\langle {\hat X_d ^{in}(\omega )\hat P_d ^{in}( - \omega )} \right\rangle \nonumber \\
&&\qquad+ \frac{{{\omega ^2} + {\Gamma ^2}/4}}{{\omega _m^2}}\left\langle {\hat X_d ^{in}(\omega )\hat X_d ^{in}( - \omega )} \right\rangle ,
 \label{noise spectrum1}
\end{eqnarray}
where the correlation functions in the Fourier space are given by
\begin{eqnarray}
&&\left\langle {\hat f(\omega )\hat f( - \omega ')} \right\rangle  = ({{\bar n}_m} + \frac{1}{2})\delta (\omega  - \omega ') \simeq \frac{{{k_B}T}}{{\hbar {\omega _m}}}\delta (\omega - \omega '), \nonumber \\
&&\left\langle {\hat X_a^{in}(\omega )\hat X_a^{in}( - \omega ')} \right\rangle  = \frac{1}{2}\left( {(2N + 1) + 2{\mathop{\rm Re}\nolimits} M} \right)\delta (\omega  - \omega '), \nonumber \\
&&\left\langle {\hat P_a^{in}(\omega )\hat P_a^{in}( - \omega ')} \right\rangle  = \frac{1}{2}\left( {(2N + 1) - 2{\mathop{\rm Re}\nolimits} M} \right)\delta (\omega  - \omega '),\nonumber \\
&&\left\langle {\hat X_a^{in}(\omega )\hat P_a^{in}( - \omega ')} \right\rangle  = \frac{i}{2}\left( {1 - 2i{\mathop{\rm Im}\nolimits} M} \right)\delta (\omega  - \omega '),\nonumber \\
&&\left\langle {\hat P_a^{in}(\omega )\hat X_a^{in}( - \omega ')} \right\rangle  = \frac{{ - i}}{2}\left( {1 + 2i{\mathop{\rm Im}\nolimits} M} \right)\delta (\omega - \omega '),\nonumber \\
&&\left\langle {\hat P_d ^{in}(\omega )\hat X_d ^{in}( - \omega ')} \right\rangle  =  - \left\langle {\hat X_d ^{in}(\omega )\hat P_d ^{in}( - \omega ')} \right\rangle  = \frac{i}{2}\delta (\omega  - \omega '),\nonumber \\
&&\left\langle {\hat X_d ^{in}(\omega )\hat X_d ^{in}( - \omega ')} \right\rangle  = \left\langle {\hat P_d ^{in}(\omega )\hat P_d ^{in}( - \omega ')} \right\rangle  = \frac{1}{2}\delta (\omega - \omega '). \nonumber \\
&&
\label{Fourier correlations}
\end{eqnarray}
Inserting these expressions one finally gets the general result of Eq. (\ref{final spectrum}).

\section{\label{App.B} Exact expression of force noise spectrum }
Based on Eq.(\ref{added force}), the exact expression of the force noise spectrum in the general case without CQNC condition is given by
\begin{eqnarray} \label{S_full}
&& S_F(\omega)= S_{th}(\omega)+ S_{f}(\omega) + S_{b}(\omega) +S_{atom}(\omega)+ S_{fb}(\omega), \nonumber \\
\end{eqnarray}
where $S_{th}(\omega)= k_B T/\hbar \omega_m$ is the thermal noise contribution, $S_{f}(\omega)$ corresponds to the field contribution, the third term is associated with the contribution of backaction noise, the fourth term corresponds to the atomic contribution, and the last term is an interference term asscoaited with the joint action of the cavity field and of the atoms. The explicit expressions are given by
\begin{subequations} \label{S_contributions}
\begin{eqnarray}
&& S_{f}(\omega)= \frac{\kappa}{g^2 \gamma_m \vert \chi_m(\omega) \vert^2} \left\{ \Delta_c  {\rm Im} \left[ Z(\omega)\left(1-2i {\rm Im}M \right) \right] \right.   \nonumber \\
&&  \left. +\left[  1+ \frac{1}{\kappa^2 \vert \chi'_a(\omega) \vert^2} -\frac{2 {\rm Re} \chi'_a(\omega)}{\kappa \vert \chi'_a(\omega) \vert^2} \right]\left(N+\frac{1}{2}-{\rm Re} M \right) \right.\nonumber \\
&& \left.  + \Delta_c^2 \vert \chi_a(\omega) \vert^2 \left(N+\frac{1}{2}+{\rm Re} M \right) \right\}  , \\
&& S_{b}(\omega)= \frac{4g^2}{\kappa \gamma_m} \left(N+\frac{1}{2}+{\rm Re} M\right) \left\vert 1+\frac{G^2}{g^2}R(\omega) \right\vert^2 , \\
&& S_{atom}(\omega)= \frac{\vert A(\omega) \vert^2}{2}  \left( 1+\frac{\omega^2 + \Gamma^2/4}{\omega_m^2} \right), \\
&& S_{fb}(\omega)=  \frac{\kappa}{\gamma_m}{\rm Im} \left[ \left(2i{\rm Im} M-1\right) \frac{Z(\omega)}{\chi_m^{\ast}(\omega)}   \left[1+\frac{G^2}{g^2}R(\omega)\right]\right] \nonumber \\
&&  -\frac{2\kappa}{\gamma_m}  \Delta_c  \vert \chi_a(\omega)  \vert^2 \left( N +\frac{1}{2} + {\rm Re}M\right) {\rm Re}\left[ \frac{1+\frac{G^2}{g^2}R(\omega)}{\chi_m^{\ast}(\omega)} \right], \nonumber  \\
\end{eqnarray}
\end{subequations}
where
\begin{subequations} \label{R,Z}
\begin{eqnarray}
&& Z(\omega)= \chi_a(\omega) \left( 1-\frac{1}{\kappa \chi'^{\ast}_a(\omega)} \right) , \\
&& R(\omega)=\frac{\chi_d(\omega)}{\chi_m(\omega)}= - \left( 1+r(\omega) \right), \\
&& r(\omega) =\frac{i \omega (\gamma_m-\Gamma)}{(\omega_m^2-\omega^2)+i\omega \Gamma} , \\
&& A(\omega)=  \frac{G}{g} \sqrt{\frac{\Gamma}{\gamma_m}} R(\omega) .
\end{eqnarray}
\end{subequations}
Notice that under perfect CQNC conditions, $1+(G^2/g^2)R(\omega)=0$, and both contributions $S_b$ and $S_{fb}$ become zero. In the Markov limit, $\kappa \gg \omega$, we keep only the zero order of $\omega/\kappa$, therefore we have
\begin{eqnarray} \label{limit}
&& \chi_a(\omega) \simeq 2/\kappa + \mathcal{O}(\omega/\kappa) , \nonumber \\
&&  \chi'^{-1}_a(\omega) \! \simeq \! \frac{\kappa}{2} \left[ \! \left( \! 1 \! + \! 4 \frac{\Delta_c^2}{\kappa^2}\right) - 4   \frac{\Delta_c}{\kappa^2} g^2 \chi_m(\omega) \left( \! 1 \! + \! \frac{G^2}{g^2} R(\omega) \right) \right], \nonumber  \\
&& Z(\omega)  \! \simeq \! \kappa^{-1} \left[ (1-4 \frac{\Delta_c^2}{\kappa^2})+4\frac{\Delta_c}{\kappa^2}\! g^2 \chi^{\ast}_m(\omega) \!  \left(1\! + \! \frac{G^2}{g^2}R^{\ast}(\omega) \right)\right], \nonumber \\
\end{eqnarray}
When we choose the optimal case of zero cavity detuning, $\Delta_c=0$, the total force noise spectrum considerably simplifies and we get
\begin{eqnarray}  \label{S_zerodetuning}
&& S(\omega)\vert_{\Delta_c=0}= \frac{k_b T}{\hbar \omega_m}+\frac{\kappa}{4g^2 \gamma_m} \frac{1}{\vert \chi_m(\omega) \vert^2}\left(N+\frac{1}{2}-{\rm Re} M \right) \nonumber \\
&&+ \frac{4g^2}{\kappa \gamma_m} \left(N+\frac{1}{2}+{\rm Re} M \right) \left\vert 1+\frac{G^2}{g^2}R(\omega) \right\vert^2 \nonumber  \\
&&+ \frac{1}{2}\left(\frac{G}{g}\right)^2 \frac{\Gamma}{\gamma_m} \vert R(\omega)\vert^2 \left( 1+\frac{\omega^2 + \Gamma^2/4}{\omega_m^2} \right) \nonumber \\
&& + {\rm Im} \left[  \frac{\left(2i{\rm Im} M-1\right)}{\gamma_m \chi_m^{\ast}(\omega)}   \left[1+\frac{G^2}{g^2}R(\omega)\right]\right] .
\end{eqnarray}
Eq. (\ref{S_zerodetuning}) shows the effect of mismatch. Without the atomic ensemble ($G=0$) the force noise spectrum of the optomechanical cavity with squeezed injection can be written as
\begin{eqnarray}\label{s_optomechanic standard with squeezing}
&& {S_{opt}^{st}}(\omega ) = \frac{{{k_B}T}}{{\hbar {\omega _m}}} + 2 {\rm Im} M  Q_m \frac{\omega_m^2-\omega^2}{\omega_m^2}   \nonumber \\
&& +{\frac{1}{4}\frac{\kappa }{{{g^2}{\gamma _m}}}\frac{1}{{{{\left| {{\chi _m}} \right|}^2}}}\left(N +\frac{1}{2}-{\rm Re}M\right)  + \frac{4{g^2}}{{\kappa {\gamma _m}}}}\! \left(N +\frac{1}{2}+{\rm Re}M\right). \nonumber \\
\end{eqnarray}
If we restrict to the injected squeezing which is optimal under perfect CQNC condition for suppressing shot noise, i.e., ${\rm {Im} M}=\phi = 0$, one can easily see how the SQL is modified by phase quadrature squeezing \cite{caves80,caves}, by minimizing over $g^2$: minimum noise is achieved at
\begin{equation}
g^2=\frac{\kappa}{4} \frac{1}{\vert \chi_m(\omega) \vert} \sqrt{\frac{2N+1-2{\rm Re} M}{2N+1+2{\rm Re} M}},
\label{g_optimum}
\end{equation}
corresponding to the modified SQL
\begin{eqnarray}\label{SQLsqueezed}
S_{SQL}=\frac{\sqrt{(2N+1)^2-4(\rm{Re}M)^2}}{\gamma_m \vert \chi_m(\omega) \vert},
\end{eqnarray}
which in the case of pure squeezed driving $\vert M \vert= \sqrt{N(N+1)}$ reproduces the usual force SQL, $S_{SQL}=1/\gamma_m \vert \chi_m(\omega) \vert$ \cite{caves80,caves}.

If in the case without atoms we do not restrict to a given squeezing phase, and we optimize not only over $g^2$ (power), but also over the phase $\phi$, we get that the optimal power is still given by Eq. (\ref{g_optimum}), and that, restricting to the pure squeezing case, the optimal squeezing phase is given by $2 {\rm Im} M = \sqrt{N(N+1)}\sin \phi = -{\rm Re}\chi_m/ {\rm Im}\chi_m$. This latter optimization allows to reach the so-called ultimate quantum limit \cite{Reynaud,pace} in the case of force sensing, which is smaller or equal (at resonance) than the SQL, and is given by
\begin{equation}\label{ultim}
   S_{ult}=\frac{|{\rm Im}\chi_m|}{\gamma_m |\chi_m(\omega) |^2}.
\end{equation}


\begin{thebibliography}{60}
\setlength{\baselineskip}{0.2 \baselineskip}
%%%%%%%%%%%%%%%%%%%%%% P1 quantum noise
\bibitem{1} C. H. Hansen, \textit{Understanding Active Noise Cancellation} (Taylor and Francis, London, 2001).
\bibitem{PRL} M. Tsang, and C.M. Caves, Phys. Rev. Lett. \textbf{105} 123601 (2010).
% P2  force measurement %%%%%%%%%%%%%%%%%%%%%%%%%%%%%%
\bibitem{PRX} M. Tsang, and C.M. Caves, Phys. Rev. X \textbf{2} 031016 (2012).
\bibitem{4} V. B. Braginsky, and F. Y. Khalili,\textit{ Quantum Measurement} edited by K. Thorne (Cambridge University Press, Cambridge, 1995).
\bibitem{5} M. Aspelmeyer, T. Kippenberg and F. Marquardt, Rev. Mod. Phys. \textbf{86}, 1391 (2014);
\bibitem{6} Y. Chen, J. Phys. B \textbf{46}, 104001 (2013).
\bibitem{7} P. Meystre, Ann. Phys. (Berlin) \textbf{525}, \textbf{215} (2013).
\bibitem{8} A. A. Clerk, M. H. Devoret, S. M. Girvin, F. Marquardt, and R. J. Schoelkopf, Rev. Mod. Phys. \textbf{82}, 1155 (2010).
\bibitem{9} S. L. Danilishin and F. Y. Khalili, Living Rev. Relat. \textbf{15}, 5 (2012).
\bibitem{10} A. Abramovici {\it et al.}, Science \textbf{256}, 325 (1992).
\bibitem{11} T. P. Purdy, R. W. Peterson, and C. A. Regal, Science \textbf{339}, 801 (2013).
\bibitem{12} K. W. Murch, K. L. Moore, S.Gupta, and D. M. Stamper-Kurn, Nat. Phys. \textbf{4}, 561 (2008).
\bibitem{13} G. M. Harry, Class. Quantum Grav. \textbf{27}, 084006 (2010); F. Acernese {\it et al.} Class. Quantum Grav. \textbf{32}, 024001 (2015).
%P3 varous proposals %%%%%%%%%%%%%%%%%%%%%%%
\bibitem{14} R. S. Bondurant and J. H. Shapiro, Phys. Rev.D \textbf{30}, 2548 (1984).
\bibitem{15} H. J. Kimble, Y. Levin, A. B. Matsko, K. S. Thorne, and S. P. Vyatchanin, Phys. Rev. D \textbf{65}, 022002 (2001).
\bibitem{16} F. Ya. Khalili, Phys. Rev. D \textbf{81}, 122002 (2010), and references therein.
\bibitem{17} R. S. Bondurant, Phys. Rev. A \textbf{34}, 3927 (1986).
\bibitem{18} T. Briant, M. Cerdonio, L. Conti, A. Heidmann, A. Lobo, and M. Pinard, Phys. Rev. D \textbf{67}, 102005 (2003).
\bibitem{19} Y. Chen, S. L. Danilishin, F. Ya. Khalili, and H. M\"uller-Ebhardt, Gen. Relat. Gravit. \textbf{43}, 671 (2010).
\bibitem{20} K. S. Thorne, R. W. P. Drever, C. M. Caves, M. Zimmermann, and V. D. Sandberg, Phys. Rev. Lett. \textbf{40}, 667 (1978).
\bibitem{21} A. A. Clerk, F. Marquardt, and K. Jacobs, New J. Phys. \textbf{10}, 095010 (2008).
\bibitem{22} J. Suh, A. J. Weinstein, C. U. Lei, E. E. Wollman, S. K. Steinke, P. Meystre, A. A. Clerk, and K. C. Schwab, Science \textbf{344}, 1262 (2015).
\bibitem{23} V. B. Braginsky, Y. I. Vorontsov, and K. S. Thorne, Science \textbf{209}, 547 (1980).
\bibitem{24} S. Chelkowski, H. Vahlbruch, B. Hage, A. Franzen, N. Lastzka, K. Danzmann, and R. Schnabel, Phys. Rev. A \textbf{71}, 013806 (2005).
\bibitem{25} C. M. Mow-Lowry, B. S. Sheard, M. B. Gray, D. E. McClelland, and S. E. Whitcomb, Phys. Rev. Lett. \textbf{92}, 161102 (2004).
\bibitem{26} B. S. Sheard, M. B. Gray, C. M. Mow-Lowry, D. E. McClelland, and S. E. Whitcomb, Phys. Rev. A \textbf{69}, 051801 (2004).
\bibitem{27} T. Caniard, P. Verlot, T. Briant, P.-F. Cohadon, and A. Heidmann, Phys. Rev. Lett. \textbf{99}, 110801 (2007).
\bibitem{28} T. Caniard {\it et al}., Phys. Rev. Lett. \textbf{99}, 110801 (2007).
\bibitem{pontin} A. Pontin, C. Biancofiore, E. Serra, A. Borrielli, F. S. Cataliotti, F. Marino, G. A. Prodi, M. Bonaldi, F. Marin, and D. Vitali, Phys. Rev. A \textbf{89}, 033810 (2014).
\bibitem{teufel}F. Lecocq, J. B. Clark, R. W. Simmonds, J. Aumentado, and J. D. Teufel, Phys. Rev. X \textbf{5}, 041037 (2015).
% P4  Caves PRL%%%%%%%%%%%%%%%%%%%%%%%%%
\bibitem{29} K. Hammerer, M. Aspelmeyer, E. S. Polzik, and P. Zoller, Phys. Rev. Lett. \textbf{102}, 020501 (2009).
\bibitem{30}E. S. Polzik and K. Hammerer, Ann. Phys. (Berlin) \textbf{527}, A15 (2015)
\bibitem{31} W. Wasilewski, K. Jensen, H. Krauter, J. J. Renema, M. V. Balabas, and E. S. Polzik, Phys. Rev. Lett. \textbf{104}, 133601 (2010).
\bibitem{maximilian} M. H. Wimmer, D. Steinmeyer, K. Hammerer, and M. Heurs, Phys. Rev. A \textbf{89}, 053836 (2014).
\bibitem{34} K. Zhang, P. Meystre, and W. Zhang, Phys. Rev. A \textbf{88}, 043632 (2013).
\bibitem{35} M. J. Woolley and A. A. Clerk, Phys. Rev. A \textbf{87}, 063846 (2013).
% P5 hybrid optomechanical setup %%%%%%%%%%%%%%%%%%%%%%%%%
\bibitem{36} P. Treutlein, C. Genes, K. Hammerer, M. Poggio, and P. Rabl, in \textit{Cavity Optomechanics}, edited by M. Aspelmeyer, T. Kippenberg, F. Marquardt (Springer, 2012).
\bibitem{37} C. Genes, H. Ritsch, and D. Vitali, Phys. Rev. A \textbf{80}, 061803(R) (2009).
\bibitem{38} K. Hammerer, K. Stannigel, C. Genes, P. Zoller, P. Treutlein, S. Camerer, D. Hunger, and T. W. Hansch, Phys. Rev. A \textbf{82}, 021803(R) (2010).
\bibitem{39} G. Ranjit, C. Montoya, and A. A. Geraci, Phys. Rev. A \textbf{91}, 013416 (2015).
\bibitem{40} S. Camerer, M. Korppi, A. Jockel, D. Hunger, T. W. Hansch, and P. Treutlein, Phys. Rev. Lett. 107, 223001 (2011).
\bibitem{41} F. Bariani, S. Singh, L. F. Buchmann, M. Vengalattore, and P. Meystre, Phys. Rev. A \textbf{90}, 062327 (2014).
\bibitem{42} A. Jockel, A. Faber, T. Kampschulte, M. Korppi, M. T. Rakher, P. Treutlein, Nat. Nanotech. \textbf{9}, 99 (2014).
\bibitem{Nori} H. Ian, Z.R. Gong, Y. Liu, C.P. Sun, and F. Nori, Phys. Rev. A \textbf{78}, 013824 (2008).
\bibitem{genes}C. Genes, D. Vitali, and P. Tombesi, Phys. Rev. A \textbf{77} 050307 (2008)
% P6 meystre  %%%%%%%%%%%%%
\bibitem{meystre} F. Bariani, H. Seok, S. Singh, M. Vengalattore, and P. Meystre, Phys. Rev. A \textbf{92}, 043817 (2015).
\bibitem{meystrecooling} F. Bariani, S. Singh, L. F. Buchmann, M. Vengalattore, and P. Meystre, Phys. Rev. A \textbf{90}, 062327 (2014).
% P7  squeexing injection %%%%%%%%%%%%%%%
\bibitem{caves80}C. Caves, Phys. Rev. Lett. \textbf{45} 75 (1980).
\bibitem{caves}C. M. Caves, Phys. Rev. D \textbf{23}, 1693 (1981).
\bibitem{Reynaud} M. T. Jaekel and S. Reynaud, Europhys. Lett. \textbf{13}, 301(1990).
\bibitem{pace}A. F. Pace, M. J. Collett, and D. F. Walls, Phys. Rev. A \textbf{47}, 3173 (1993).
\bibitem{QuantumMeasurement} V. B. Braginsky and F. Ya. Khalili, \textit{Quantum Measurement } (Cambridge, University Press,1992).
\bibitem{McKenzie} K. McKenzie, D. A. Shaddock, D. E. McClelland, B. C. Buchler, and P. K. Lam, Phys. Rev. Lett. \textbf{88}, 231102 (2002).
\bibitem{Chen} Y. Chen, J. Phys. B.: At. Mol. Opt. Phys. \textbf{46}, 104001 (2013).
\bibitem{LIGO} The LIGO Scientific Collaboration, Nature Photonics 7, 613 (2013).
\bibitem{Mach-Zehnder} Min Xiao, Ling-Au Wu, and H. J. Kimble, Phys. Rev. Lett \textbf{59}, 278 (1987).
\bibitem{Sagnac} T. Eberle, S. Steinlechner, J. Bauchrowitz, V. H\"andchen, H. Vahlbruch, M. Mehmet, H. M\"uller-Ebhardt, and R. Schnabel, Phys. Rev. Lett. \textbf{104}, 251102 (2010).
\bibitem{polarization} P. Grangier, R. E. Slusher, B. Yurke, and A. LaPorta, Phys. Rev. Lett \textbf{59}, 2153 (1987).
\bibitem{transduction} U. B. Hoff, G. I. Harris, L. S. Madsen, H. Kerdoncuff, M. Lassen, B. M. Nielsen, W. P. Bowen, and U. L. Andersen, Opt. Lett. \textbf{38}, 1413 (2013)
\bibitem{phase-squeezed state} K. Iwasawa, K. Makino, H. Yonezawa, M. Tsang, A. Davidovic, E. Huntington, and A. Furusawa, Phys. Rev. Lett. \textbf{111}, 163602 (2013).
\bibitem{Clark}J. B. Clark, F. Lecocq, R. W. Simmonds, J. Aumentado, and J. D. Teufel, arXiv:1601.02689.
\bibitem{intracavity squeezing} V. Peano, H. G. L. Schwefel, Ch. Marquardt, and F. Marquardt, Phys. Rev. Lett. \textbf{115}, 243603 (2015).
\bibitem{Lotfipor} H. Lotfipour, S. Shahidani, R. Roknizadeh, and M. H. Naderi, Phys. Rev. A \textbf{93}, 053827 (2016).
% single mode %%%%%%%%%%%%%%%%%%%%%%%%%%%%
\bibitem{53} C. K. Law, Phys. Rev. A \textbf{51}, 2537 (1995).
\bibitem{54} C. Genes, D. Vitali, and P. Tombesi, New J. Phys. \textbf{10}, 095009 (2008).
% effective two level atom and Faradey  %%%%%%%%%%%%%%%%%%%%%%
\bibitem{Quantum interface between light and atomic ensembles} K. Hammerer, A. S. Sorensen, and E. S. Polzik, Rev. Mod. Phys. \textbf{82}, 1041 (2010).
\bibitem{Atomic2} S. Chaudhury, G. A. Smith, K. Schulz, and P. S. Jessen, Phys. Rev. Lett. \textbf{96}, 043001(2006).
\bibitem{tomography} I. H. Deutsch and P. S. Jessen, Opt. Commun. \textbf{283}, 681(2010).
\bibitem{magnetometry} M. Napolitano, M. Koschorreck, B. Dubost, N. Behbood, R. J. Sewell, and M. W. Mitchell, Nature (London) \textbf{471},486 (2011).
%%%%%%%%%%%%%%%%%%%%%%%%%%%%
\bibitem{HP}T. Holstein and H. Primakoff, Phys. Rev. \textbf{58}, 1098 (1940).
\bibitem{vitali noise membrane} V. Giovannetti and D. Vitali, Phys. Rev. A \textbf{63}, 023812 (2001).
\bibitem{GZ}C. W. Gardiner and P. Zoller, {\it Quantum Noise} (Springer, Berlin, 2000).
\bibitem{55} K. Jahne, C. Genes, K. Hammerer, M. Wallquist, E. S. Polzik, and P. Zoller, Phys. Rev. A \textbf{79}, 063819 (2009)
\bibitem{56} M. J. Collett and C. W. Gardiner, Phys. Rev. A \textbf{30}, 1386 (1984).
\bibitem{input-output} D. F. Walls, and G. J. Milburn, \textit{Quantum Optics}, 2nd edition (Springer, Berlin, 2008).
\bibitem{heinze}G. Heinze, C. Hubrich, and T. Halfmann, Phys. Rev.
Lett. \textbf{111}, 033601 (2013).
\bibitem{dudin}Y. O. Dudin, L. Li, and A. Kuzmich, Phys. Rev. A \textbf{87}, 031801(R) (2013).
\bibitem{thompson}J. D. Thompson, B. M. Zwickl, A. M Jayich, F. Marquardt, S. M. Girvin, J. G. E. Harris, Nature (London) \textbf{452}, 72 (2008).
\end{thebibliography}
\end{document}